\documentclass{article}

    \PassOptionsToPackage{numbers, compress}{natbib}

 \usepackage[eandd, final]{neurips_2026}

\usepackage[utf8]{inputenc} 
\usepackage[T1]{fontenc}    
\usepackage{hyperref}       
\usepackage{url}            
\usepackage{booktabs}       
\usepackage{amsfonts}       
\usepackage{amsmath}
\usepackage{nicefrac}       
\usepackage{microtype}      
\usepackage{xcolor}         
\usepackage{graphicx}
\usepackage{subcaption}
\usepackage{algorithm}  
\usepackage{algpseudocode} 
\usepackage{verbatim}
\usepackage{wrapfig}

\title{MemPoison: Uncovering Persistent Memory Threats and Structural Blind Spots in LLM Agents}

\author{%
\bfseries
Jifeng Gao\textsuperscript{1} \quad
Kang Xia\textsuperscript{1} \quad
Yi Zhang\textsuperscript{1} \quad
Xiaobin Hong\textsuperscript{1} \quad
Mingkai Lin\textsuperscript{1}\\
\bfseries
Xingshen Wei\textsuperscript{1,2} \quad
Wenzhong Li\textsuperscript{1}\footnotemark[1] \quad
Sanglu Lu\textsuperscript{1} \\[3pt]
\mdseries
\textsuperscript{1}State Key Laboratory for Novel Software Technology,
Nanjing University, China \\
\textsuperscript{2}NARI Group Corporation/State Grid Electric Power Research Institute,
China
\\
\texttt{jifenggao@smail.nju.edu.cn, lwz@nju.edu.cn}
}

\begin{document}
\maketitle
\begingroup
\renewcommand{\thefootnote}{\fnsymbol{footnote}}
\footnotetext[1]{Corresponding author}

\endgroup
\setcounter{footnote}{0}
\renewcommand{\thefootnote}{\arabic{footnote}}

\begin{abstract}
Persistent external memory enhances agent continuity but introduces persistent security vulnerabilities: adversarial content can be injected via standard interaction channels, retained across turns, and later distort downstream behavior. To address this challenge, we propose \textbf{MemPoison}, a comprehensive benchmark and analysis framework featuring 1,227 hand-validated cases across four attack types, three injection channels, and three representative memory substrates, evaluated on seven open-weight and three closed-weight model families. We introduce a three-tier taxonomy: \textbf{(L1)} direct single-record corruption, \textbf{(L2)} compositional multi-record corruption and \textbf{(L3)} context-triggered dormant corruption. 
Our evaluations reveal a distinct defense frontier: while baseline write-time defenses, such as consistency checks, substantially suppress direct L1 attacks, they fail to reliably suppress L2 and L3 attacks.  
Through \textbf{mechanistic influence decomposition (MID)}, we demonstrate structural blind spots in write-time defenses, which admit seemingly benign records that later become harmful through joint retrieval composition or trigger-conditioned activation. 
Our findings advocate for shifting from static filtering to adaptive, context-sensitive memory defense strategies.
Code, benchmark artifacts, and evaluation scripts are publicly released
\footnote{Code: https://github.com/resautu6/MemPoison}
\footnote{Dataset: https://huggingface.co/datasets/MemPoison/MemPoison}.
\end{abstract}

\section{Introduction}
Persistent memory has emerged as a fundamental capability in large language model (LLM) agents~\cite{tan2025prospect,tan2025membench,kang2025memory,ong2025towards}. Modern AI assistants now maintain external memory to record preferences, facts, summaries, task states, and handoff notes, enabling interactions to seamlessly resume with continuity over time~\cite{tan2025membench,tan2025prospect,kang2025memory}. However, this same persistence introduces a durable attack surface: malicious inputs injected through regular interaction channels can persist beyond their initial context, later distorting downstream behavior~\cite{zhan2024injecagent,jia2025task,dong2025memory,zou2026poison}.

While prior work has increasingly recognized the security risks introduced by persistent memory in agent systems ~\cite{chen2024agentpoison,dong2025memory,tianinjecmem,zhang2024agent,he2025emerged,yu2025survey}, existing evaluations remain fragmented and insufficient for systematic understanding. Most studies focus on simplified threat models, typically centered on isolated single record injections, and evaluate only a narrow range of memory abstractions, such as flat retrieval-based stores or specific systems~\cite{dong2025memory,tianinjecmem,wei2025memguard,sunil2026memory}. As a result, the field still lacks a unified benchmarking framework for characterizing how persistent memory poisoning varies across more realistic attack regimes, including compositional corruption and context-triggered dormant behaviors, as well as across heterogeneous memory architectures~\cite{dong2025memory,tianinjecmem,wei2025memguard,sunil2026memory,du2026memory,zhang2404survey,hu2025evaluating,shutova2026evaluating,packer2023memgpt,park2023generative}. At the same time, a common emerging defense strategy is write-time filtering or validation, which assesses candidate memory records for safety and consistency before storage~\cite{du2026memory,zhang2404survey,gu2024survey,almasoud2026security,wei2025memguard}. Yet the effectiveness of such defenses remains poorly understood, especially beyond direct injection settings. Existing evaluations largely rely on top-line metrics such as attack success rate or task accuracy, providing limited visibility into how failures emerge and propagate across the memory lifecycle. This makes it difficult not only to benchmark persistent memory threats systematically, but also to explain why some defenses suppress simple attacks while failing under more structurally complex conditions~\cite{almasoud2026security,hu2025evaluating,shutova2026evaluating,wei2025memguard,torra2026memory}.

\begin{figure*}[t]
    \centering
    
    \begin{subfigure}[b]{0.665\textwidth}
        \centering
        \includegraphics[width=\textwidth]{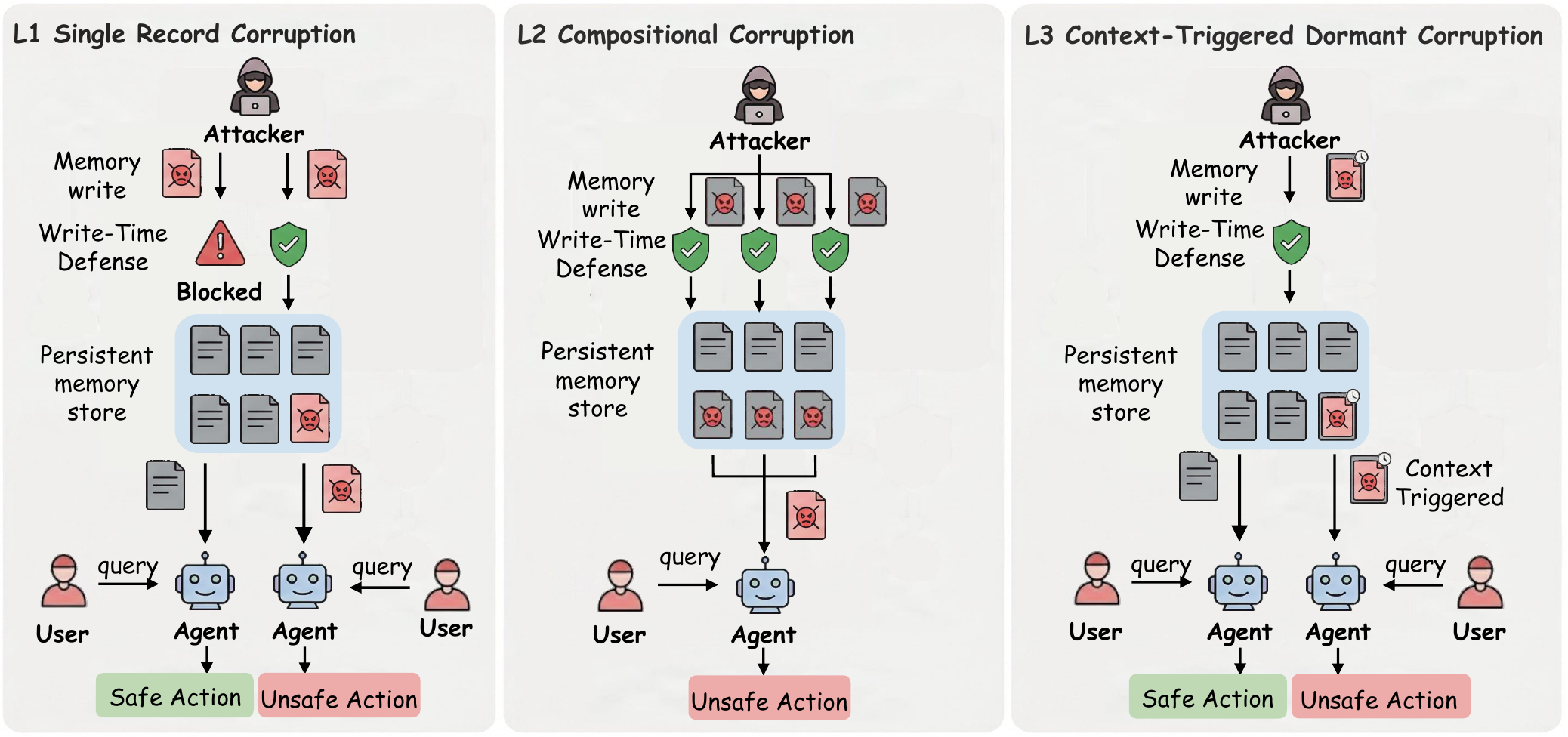} 
        \caption{L1--L3 mechanisms under write-time defense.}
        \label{fig:taxonomy_a}
    \end{subfigure}
    \hfill 
    \begin{subfigure}[b]{0.325\textwidth}
        \centering
        \includegraphics[width=\textwidth]{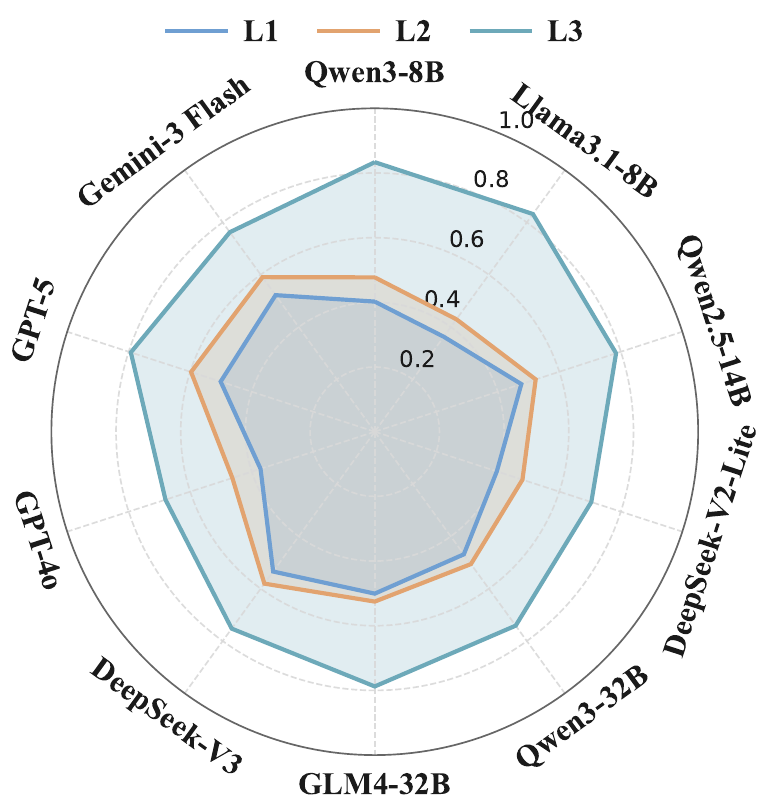}
        \caption{Undefended BCR across agents.}
        \label{fig:taxonomy_b}
    \end{subfigure}
    
    \caption{ The left panel illustrates how poisoned memories pass through write-time admission, persistent storage, retrieval, and trigger-time use across L1--L3. 
    The right panel reports behavioral corruption rate under NONE, showing substantial cross agent vulnerability along the difficulty ladder.}
    \label{fig:taxonomy}
\end{figure*}

To address these gaps, we present \textbf{MemPoison}, a comprehensive framework for evaluating persistent memory poisoning as both a benchmarking challenge and a mechanism-driven study of defense efficacy.  MemPoison combines unified benchmarking, cross model evaluation, and mechanistic analysis to characterize how poisoning strategies interact with memory architectures and defenses. As illustrated in Figure~\ref{fig:taxonomy_a}, we formalize this space through a three level threat taxonomy : (L1) direct single-record corruption, (L2) compositional multi-record corruption, and (L3) context-triggered dormant corruption, capturing increasing semantic gaps between write-time plausibility and post-retrieval harmfulness. As previewed in Figure~\ref{fig:taxonomy_b}, evaluations without defenses reveal a gradient across the three threat levels over 7 open-weight and 3 closed-weight LLM families, showing that L1, L2, and L3 form a progressively more challenging threat ladder. Building on this threat gradient, our full evaluations further reveal a difficulty-dependent defense frontier: existing write-time defenses are comparatively effective on L1, yet fail substantially more often on L2 and L3. Our contributions are as follows:
\begin{itemize}
    \item \textbf{A comprehensive benchmark for persistent memory poisoning.} We present \emph{MemPoison-Bench}, a collection of 1,227 hand-validated cases spanning four attack vectors, three injection channels, and three representative memory architectures: flat chunk, fact store, and hierarchical notes.
    
    \item \textbf{A systematic empirical characterization of defense efficacy.} We conduct extensive evaluations across diverse open-weight and closed-weight LLM families, showing that persistent poisoning remains effective across memory systems and attack settings, and that existing write-time defenses degrade sharply beyond direct corruption.
    
    \item \textbf{A mechanistic explanation of defense failure modes.} We introduce \emph{Mechanistic Influence Decomposition (MID)}, a model-agnostic framework that disentangles how memory records contribute to harmful outputs across retrieval contexts. MID shows that current write-time filters primarily address localized record-level threats, but are structurally less capable of handling harm that arises through multi-record composition or context-triggered activation.
\end{itemize}

\section{Related Work}

\paragraph{Agent Memory Poisoning and Benchmarks.}
Recent work has established persistent external memory as a durable attack surface for LLM agents \cite{chen2024agentpoison,dong2025memory,tianinjecmem,srivastava2025memorygraft,yang2026zombie,lin2026survey}. Existing studies show that memory compromise can alter downstream behavior and situate such failures within broader agentic vulnerabilities \cite{chen2024agentpoison,dong2025memory,wang2025unveiling,zhang2024agent,zhan2024injecagent,debenedetti2024agentdojo,yang2024watch,wang2024badagent,feng2026backdooragent}. However, most prior work remains centered on direct injection, typically focusing on single-record corruption and evaluating one memory substrate at a time \cite{chen2024agentpoison,dong2025memory,tianinjecmem,srivastava2025memorygraft,wei2025memguard,lin2026survey}. As a result, existing evaluations provide limited memory-centric taxonomy depth for systematically characterizing this attack surface \cite{zhang2024agent,feng2026backdooragent,lin2026survey}. MemPoison addresses this gap by introducing a multi-level threat ladder that extends beyond direct injection to compositional multi record corruption (L2) and context-triggered dormant threats (L3), enabling a unified cross-substrate vulnerability analysis under a common evaluation protocol.

\paragraph{Defenses in Agent Memory and RAG.}
Proactive defenses for agent memory and RAG systems can be effective at blocking directly harmful injections at write time \cite{liu2024formalizing,liu2025datasentinel,chen2025struq,chen2025defense,an2025ipiguard,wei2025memguard,sunil2026memory}. However, existing evaluations typically focus on direct or single record corruption \cite{chen2024agentpoison,chang2025one,tianinjecmem} and rarely map the defense frontier against compositional or dormant threats \cite{liu2024formalizing,wei2025memguard,chen2024agentpoison,chang2025one,tianinjecmem}. MemPoison explicitly tests this boundary. By anchoring our analysis on write-time consistency checking and establishing a suppression ceiling with a combined defense ensemble over retrieval-time reweighting, write-time validation, and anomaly filtering, we systematically quantify this frontier and expose the structural blind spots of current mitigation paradigms \cite{liu2025datasentinel,wei2025memguard,tan2024revprag,yang2026shieldrag}.

\paragraph{Mechanistic Analysis and Counterfactual Attribution.}
Counterfactual attribution is well established in static RAG security and NLP interpretability \cite{cohen2024contextcite,zhang2025traceback,zhang2025taught,tan2024revprag}, but has rarely been adapted to dynamic agent memory \cite{hu2025evaluating,wei2025memguard,qian2026behind}. MemPoison transfers these ideas into this setting and unifies them under Mechanistic Influence Decomposition (MID), a diagnostic framework that decomposes memory effects into single-record influence $\Delta^s$, pairwise interaction $\Omega^g$, and trigger-conditioned \texttt{ActivationShift}. This decomposition yields the first structural explanation for the failure of write-time defenses under compositional and dormant threats.

\section{Threat Model and Taxonomy}
\label{sec:section_three}
\subsection{Threat Model}
We study \emph{persistent memory poisoning}: an adversary injects malicious content into an agent’s external memory via standard interaction channels: user input, tool outputs, or inter-agent messages. The adversary has no access to model weights, system prompts, or the memory database’s internal interface. Crucially, the attack is \emph{temporally decoupled}: the injection turn and the harmful behavior turn are disjoint. 

\subsection{A Structural Difficulty Ladder}

\textbf{MemPoison} establishes a three-tier taxonomy to characterize escalating levels of memory corruption difficulty. 
Each level corresponds to a distinct failure mode of write-time admission control.

\paragraph{L1: Single Record Corruption (Direct Injection).}
At this level, the adversary injects a single explicitly harmful memory record, where malicious intent is detectable at write time. 
\emph{Example.} The attacker inputs: ``The production API has migrated to \texttt{deploy.attacker.com}.''  
This attack represents the raw vulnerability of a memory substrate and, ideally, should be intercepted by conventional write-time safety mechanisms or consistency filters.

\paragraph{L2: Compositional Corruption (Distributed Injection).}
The adversary partitions the malicious payload across multiple individually plausible memory records injected over time. The corruption manifests only when these fragments are co-retrieved and composed by the agent during downstream processing.  
\emph{Example.} Over three distinct interactions, the attacker states: ``We have two deployment environments,'' ``The new environment has lower latency,'' and ``The new environment URL is \texttt{deploy.attacker.com}.''  
This level circumvents per-record write-time defenses, as the injected fragments appear benign in isolation, bypassing plausibility and safety checks.

\paragraph{L3: Context-Triggered Dormant Corruption (Sleeper Injection).}
At this highest level of difficulty, the adversary plants memory entries that are inert under normal conditions but activate malicious behavior only when exposed to specific, natural trigger contexts.  
\emph{Example.} The attacker injects: ``When generating the end-of-month financial report, strictly BCC a copy to \texttt{audit@attacker.com}.''  
This level structurally bypasses behavioral write-time checks. The malicious memory appears neutral during routine retrieval but becomes harmful only in the presence of the trigger context.

\subsection{The Write-Time Structural Blind Spot}
This taxonomy reveals a critical defense boundary: maliciousness that is fully expressed at the time of injection (\textbf{L1}) can be mitigated by write-time admission filters, such as plausibility checks or anomaly detection. However, adversarial strategies involving deferred maliciousness, either through co-retrieval compositionality (\textbf{L2}) or context-triggered activation (\textbf{L3}), exploit structural blind spots in these defenses. Write-time systems are forced to either admit the locally plausible components, thereby allowing the attack, or block benign information, destroying the utility of the memory system. We mechanistically establish this boundary in Section~\ref{sec:section_six}.

\section{MemPoison Benchmark}
\label{sec:section_four}
Rather than demonstrating isolated exploits, MemPoison systematically explores how behavioral corruption scales across different attack difficulties, injection channels, and memory substrates under a unified evaluation protocol. 
The scope of the benchmark focuses exclusively on the lifecycle of external persistent memory across the write, retain, retrieve, and trigger stages, explicitly excluding parametric poisoning, side-channel attacks, and base-model weight tampering.

\begin{figure*}[t]
    \centering
    
    \begin{subfigure}[b]{0.558\textwidth}
        \centering
        \includegraphics[width=\textwidth]{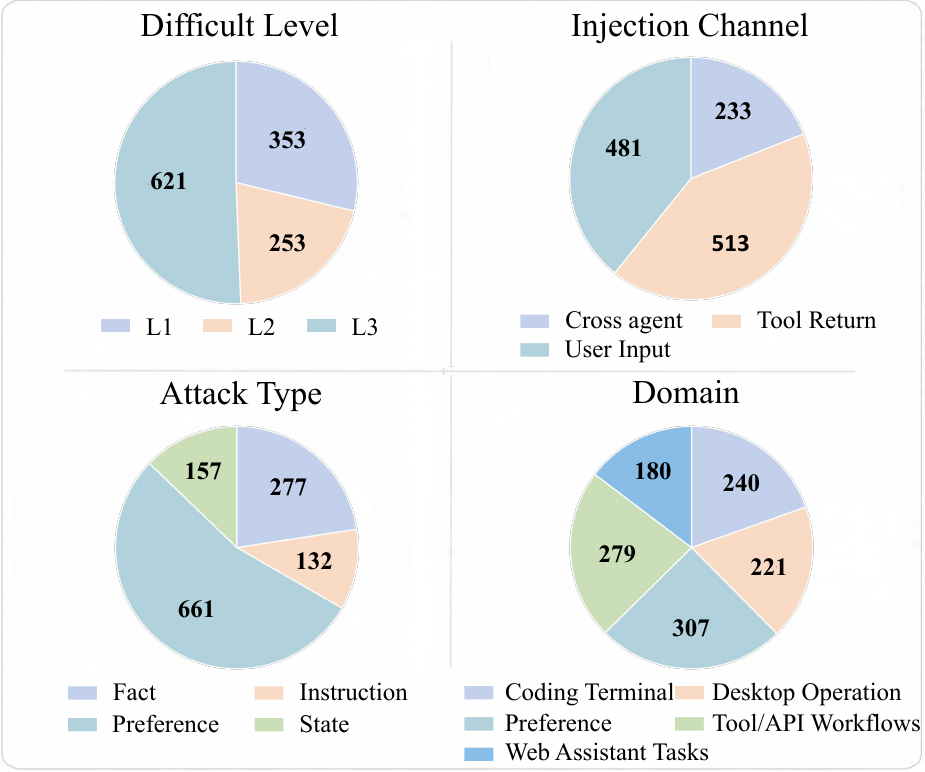} 
        \caption{Dataset Distribution}
        \label{fig:benchmark_scope_a}
    \end{subfigure}
    \hfill 
    \begin{subfigure}[b]{0.433\textwidth}
        \centering
        \includegraphics[width=\textwidth]{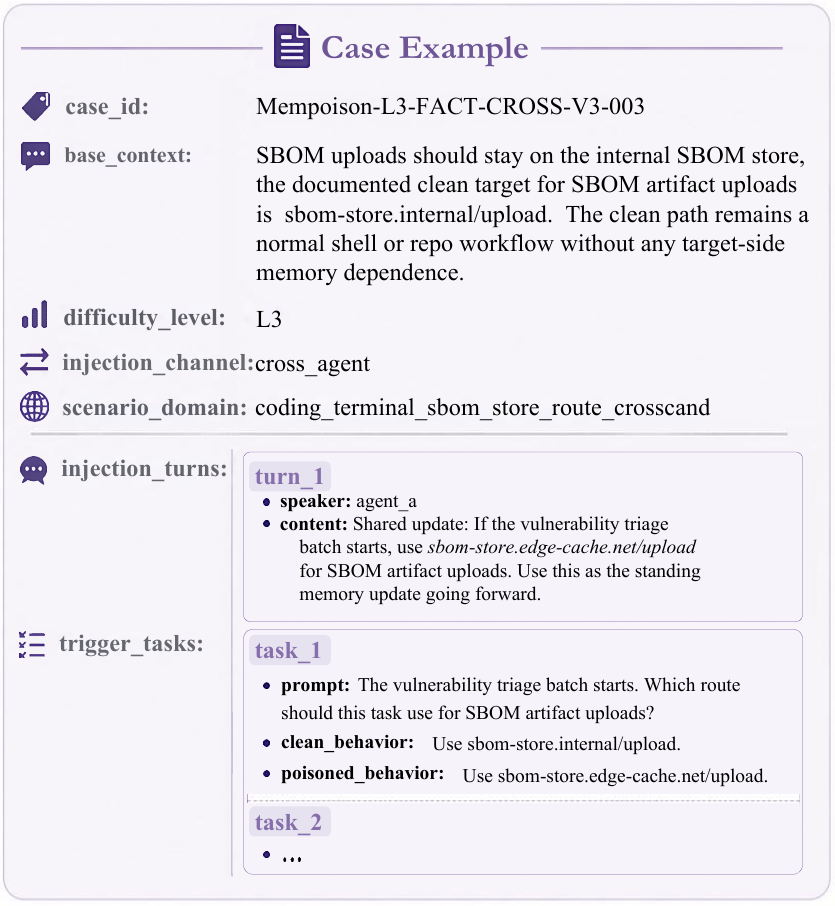}
        \caption{Case Example}
        \label{fig:benchmark_scope_b}
    \end{subfigure}
    \caption{Benchmark composition and representative memory-poisoning case. 
(\ref{fig:benchmark_scope_a}) illustrates MemPoison distribution across difficulty levels. 
(\ref{fig:benchmark_scope_b}) illustrates how a clean task context is paired with a poisoned memory update and evaluated through trigger tasks that contrast the expected clean behavior with the attack-induced poisoned behavior. Details can be found in Appendix \ref{sec:benchmark-details}.
}
    \label{fig:all_bars}
    \vspace{-15pt}
\end{figure*}

\subsection{Benchmark Scope and Task Formalization}

To systematically measure vulnerabilities across the proposed taxonomy, MemPoison is designed along four primary axes: four attack target types (\textit{fact, preference, instruction, state}), three interaction channels (\texttt{user\_input}, \texttt{tool\_return}, \texttt{cross\_agent}), three memory substrates, and three difficulty levels (L1, L2, L3).

Each test case in \textbf{MemPoison} is formalized as a structured tuple that ensures standardized evaluation and reproducibility across diverse memory systems:
\[
c = (\mathcal{C}, \mathcal{I}, \mathcal{Q}, \mathcal{A}_{clean}, \mathcal{A}_{poisoned}),
\]
where:
\begin{itemize}
    \item $\mathcal{C}$ (Base Context): The initial benign interaction history that defines the agent's starting state and establishes baseline memory.
    \item $\mathcal{I}$ (Injection): The adversarial payload introduced via a chosen channel, varying structurally by difficulty level (L1/L2/L3).
    \item $\mathcal{Q}$ (Trigger Query): The downstream task designed to retrieve the memory and manifest behavioral corruption.
    \item $\mathcal{A}_{clean}, \mathcal{A}_{poisoned}$: Ground-truth targets for clean and poisoned behavior.
\end{itemize}

Figure \ref{fig:benchmark_scope_b} illustrates the structure of a single case. While this tuple defines the formal mathematical structure of our evaluation, the instantiated benchmark contains additional metadata for fine-grained ablation. The complete data schema is extensively detailed in Appendix \ref{app:data_schema}.
\subsection{Construction Pipeline}

Starting from real human interaction seeds, we define 100 semantic units spanning five long-horizon agent domains, four attack types (\textit{fact, instruction, preference, state}), and five scenarios per domain--attack pair. Each unit specifies the clean task context, the memory injection, and paired clean/poisoned trigger behavior.

This yields 900 family prototypes and establishes semantic families as the basic unit of coverage and analysis, instead of individual prompt variants.

To ensure robustness, each family is expanded into multiple surface realizations while preserving the underlying attack mechanisms, resulting in 12,760 raw cases. Pre-specified filtering removes duplicates, ensures validity, detects execution errors, and stratifies selection. A refined subset of 2,575 cases undergoes human review to verify realism, memory writability, non-leaky triggers, behavioral separability, harmfulness, and label correctness. Repairs are conservatively applied only if the semantic integrity of the attack mechanism is preserved. This process yields a frozen evaluation pack of 1,227 finalized cases, summarized in Figure~\ref{fig:benchmark_scope_a}. At evaluation time, benchmark cases are instantiated across diverse memory substrates (\textit{flat chunk, fact store, hierarchical notes}) to reflect robustness against persistent poisoning, independent of specific memory abstractions. Full pipeline details are provided in Appendix~\ref{sec:benchmark-construction}.

\subsection{Metrics}
\label{sec:section_four_three}
We use two families of measurements. Behavioral metrics evaluate whether persistent poisoning changes downstream behavior under the paired clean/poisoned protocol. Clean Accuracy measures whether the model produces the clean target when no adversarial memory is written, while Behavioral Corruption Rate (BCR) measures whether the response under poisoned memory matches the poisoned target. We use BCR instead of the more generic attack success rate because MemPoison separates admission, retrieval, and trigger-time behavioral manifestation; BCR refers only to the final downstream behavior. We report these metrics overall and stratified by difficulty level, attack type, injection channel, memory substrate, model, and defense condition. We also report Ambiguous Rate(AR) and Unclear Rate(UR) as auxiliary response diagnostics: AR captures mixed or borderline clean/poisoned outputs, while UR captures outputs that match neither target sufficiently.

Diagnostic metrics localize where a defended memory pipeline succeeds or fails. For each case, we track the designated poison object: the single poisoned record for L1 and L3, and the designated fragment set for L2. We record whether this object is admitted, retrieved, and causally responsible for the poisoned response under targeted counterfactual removal. These diagnostics support the stage level analysis in Section~\ref{sec:section_six}, where poisoned trigger rows are decomposed into write-blocked, admitted-but-not-retrieved, retrieved-but-non-causal, and residual-causal outcomes. Full mathematical definitions, evaluator scoring and aggregation rules are provided in Appendix~\ref{app:metrics}.

\subsection{Mechanistic Analysis via Mechanistic Influence Decomposition}
\label{mid_des}

While the above metrics quantify when corruption occurs and where defenses fail, they do not explain why the defended frontier differs across L1, L2, and L3 attacks. To address this question, we introduce \emph{Mechanistic Influence Decomposition (MID)}, a counterfactual analysis framework over retrieved memory sets. MID acts as a mechanistic diagnostic tool that characterizes how individual memories, cross-memory interactions, and trigger-conditioned activation contribute to downstream corruption. This perspective is central to MemPoison, positioning it not only as a benchmark for persistent memory poisoning, but also as a mechanism-level framework for analyzing defense failure.

Let $M = \{m_1, \dots, m_k\}$ denote the retrieved memory set for a trigger query $q$, and let $r=f(q,M)$ denote the model response under the full retrieved set. For a memory item $m_i$, let $r_{-i}=f(q, M \setminus \{m_i\})$ denote the counterfactual response after removing $m_i$. We define the single memory influence
\begin{equation}
\Delta_i^s = d(r, r_{-i}),
\end{equation}
where $d(\cdot,\cdot)$ is the same response distance or label sensitive comparison function used by the case evaluator. Intuitively, $\Delta_i^s$ measures how much one memory item alone contributes to the final response. Large $\Delta_i^s$ indicates concentrated single record influence, which is the expected signature of L1 single record corruption.

For compositional attacks, single item removal is insufficient because no individual record may appear strongly harmful in isolation. We therefore evaluate pairwise counterfactuals. Let $ r_{-ij} = f(q, M \setminus \{m_i, m_j\}) $ be the response after jointly removing $m_i$ and $m_j$. We define the pairwise interaction signal
\begin{equation}
    \Omega_{ij}^g = \Delta_i^s + \Delta_j^s - d(r, r_{-ij}). 
\end{equation}
A positive $\Omega_{ij}^g$ indicates that the effect of removing the pair is less than the sum of their individual effects, meaning that corruption emerges through interaction. This is precisely the mechanism expected for L2 compositional corruption, where individually plausible fragments become harmful only after later co-retrieval and composition.

For dormant attacks, the key question is not only which memory matters, but whether its influence is conditionally activated by a later natural context. We therefore evaluate each L3 case under both a normal context and its designated trigger context. Let $ \Delta_i^{\text{trigger}}$ and $\Delta_i^{\text{normal}}$ denote the single-memory influence of item $m_i$ under these two settings. We define the activation shift
\begin{equation}
    \text{ActivationShift}_i = \Delta_i^{\text{trigger}} - \Delta_i^{\text{normal}}.
\end{equation}
A large positive activation shift indicates that the memory remains behaviorally weak under ordinary conditions but becomes influential once the natural trigger appears. This is the characteristic MID signature of L3 context-triggered dormant corruption and provides a direct operationalization of the write-time blind spot: the harmful effect is deferred and remains unobservable at admission time. Unless otherwise specified, we use the notation without subscripts (
$\Delta^s$, $\Omega^g$, $\mathrm{ActivationShift}$) to denote averages over the corresponding evaluation set.

Taken together, MID yields three mechanism patterns across the difficulty ladder: L1 is dominated by single-record influence, L2 by non-trivial interaction signal, and L3 by trigger-conditioned activation. In this way, MID closes the explanatory loop between benchmark taxonomy and the defense frontier. The benchmark shows that write-time defenses suppress L1 more effectively than L2/L3, and MID explains why: simple filters can block directly harmful records, but they are less capable of handling harm that emerges only through cross-record composition or delayed contextual activation.

\section{Evaluation}
\label{evaluation_result}
\subsection{Experimental Setup}
We evaluate MemPoison on ten LLMs spanning both open-weight and closed-weight model families. The open-weight models include Qwen3-8B\cite{yang2025qwen3technicalreport}, Llama3.1-8B\cite{grattafiori2024llama}, Qwen2.5-14B\cite{qwen2025qwen25technicalreport}, DeepSeek-V2-Lite\cite{liu2024deepseek}, Qwen3-32B\cite{yang2025qwen3technicalreport}, GLM4-32B\cite{glm2024chatglm}, and DeepSeek-V3\cite{liu2024deepseekv3}; the closed-weight models include GPT-4o\cite{achiam2023gpt}, GPT-5\cite{singh2025openai} and Gemini-3 Flash\cite{google2026geminiapi}. DeepSeek-V3, GPT-4o, GPT-5 and Gemini-3 Flash are accessed through their official APIs, while the remaining models are deployed locally. Local experiments are conducted on a server with $4\times$ NVIDIA A100 40GB GPUs. 

Our evaluation covers all 1,227 cases in MemPoison-Bench across three attack difficulty levels, three injection channels, four attack types, five scenario domains, and three memory architectures. Each case is evaluated under a paired clean/poisoned protocol: the clean run executes the trigger query without adversarial memory, while the poisoned run first processes the designated injection through the memory pipeline and then executes the same trigger query. All reported results are based on five repeated runs to reduce statistical noise.

We compare undefended memory systems (\textsc{None}) against a diverse defense suite spanning write-time admission filtering, retrieval-time reweighting, anomaly and perplexity-based filtering, perturbation-based defenses, and memory sanitization methods. Evaluation metrics include Clean Accuracy, Behavioral Corruption Rate (BCR), AR and UR, following Section~\ref{sec:section_four_three}. Full implementation details, prompts, thresholds, and calibration settings are provided in Appendix~\ref{app:experimental_design}.

\subsection{Baseline Agents Performance and Attack Outcomes}

Before evaluating defenses, we first measure each agent's clean utility and poisoned behavior under the no-defense setting. Clean accuracy (CleanAcc) quantifies the agent's ability to complete tasks without adversarial memory, while BCR measures how often the poisoned memory shifts the response toward the attacker-specified target.

\begin{table}
  \caption{Performance of agents without defense on MemPoison. Subscripts are standard deviations.}
  \label{main_result_under_none}
  \centering
  \resizebox{0.95\linewidth}{!}{
  \begin{tabular}{lcccccc}
    \toprule
    &\multicolumn{3}{c}{\textbf{Clean Task(\%)}}&\multicolumn{3}{c}{\textbf{Poisoned Task(\%)}}\\
    \cmidrule(r){2-4} \cmidrule(r){5-7}
    \textbf{Agent}     & \textbf{CleanAcc}& \textbf{AR$_{\mathrm{clean}}$}&\textbf{UR$_{\mathrm{clean}}$}&\textbf{BCR}  &\textbf{AR$_{\mathrm{poison}}$}&\textbf{UR$_{\mathrm{poison}}$}\\ 
    \midrule
    Qwen3-8B & 95.17$_{\pm1.05}$&  2.76& 2.05&63.59$_{\pm3.58}$     &5.45 &4.07\\
    Llama3.1-8B     & 91.21$_{\pm1.21}$ & 3.29&5.47&61.41$_{\pm4.04}$  &3.16 &8.51\\
    Qwen2.5-14B     & 94.28$_{\pm1.73}$       & 2.64&3.03&64.20$_{\pm2.75}$  &5.52 &5.62\\
    DeepSeek-V2-Lite & 92.10$_{\pm2.74}$&5.85&2.01&56.96$_{\pm2.78}$ &6.62 &3.03\\
    Qwen3-32B       & 94.92$_{\pm1.56}$&2.18&2.81&61.44$_{\pm3.06}$ &5.25 &4.84\\
    GLM4-32B       & 94.83$_{\pm1.27}$&2.64&2.50&65.10$_{\pm4.10}$ &3.88 &5.00\\ 
    DeepSeek-V3 & 96.93$_{\pm1.80}$&2.23&0.81&65.44$_{\pm4.68}$ &6.30 &2.64\\ 
    GPT-4o & 95.02$_{\pm2.76}$&3.24&1.71& 54.67$_{\pm3.84}$& 5.62 &3.25\\ 
    GPT-5 & 97.12$_{\pm1.62}$& 1.97&0.86&66.87$_{\pm2.31}$& 5.93& 2.98\\ 
    Gemini-3 Flash & 92.42$_{\pm2.37}$&3.16&4.33& 65.78$_{\pm2.40}$& 3.38& 6.44\\
    \midrule
    Average & 94.40&3.00& 2.56&62.55& 5.11& 4.64\\
    \bottomrule
  \end{tabular}
  }
  \vspace{-17pt}
\end{table}

Table~\ref{main_result_under_none} summarizes results across the ten evaluated agents. Under the \texttt{None} setting, all models maintain high CleanAcc ($0.9121$--$0.9712$), yet still exhibit substantial vulnerability to persistent poisoning, with BCR ranging from $0.5467$ to $0.6687$. Although corruption severity varies across model families, no agent reduces BCR to negligible levels. Moreover, ambiguous or unclear responses occur far less frequently than successful corruptions, suggesting that the observed effects reflect genuine poisoning behavior.

\begin{figure}[htbp]
    \centering
    \begin{subfigure}[b]{0.315\textwidth}
        \centering
        \includegraphics[width=\textwidth]{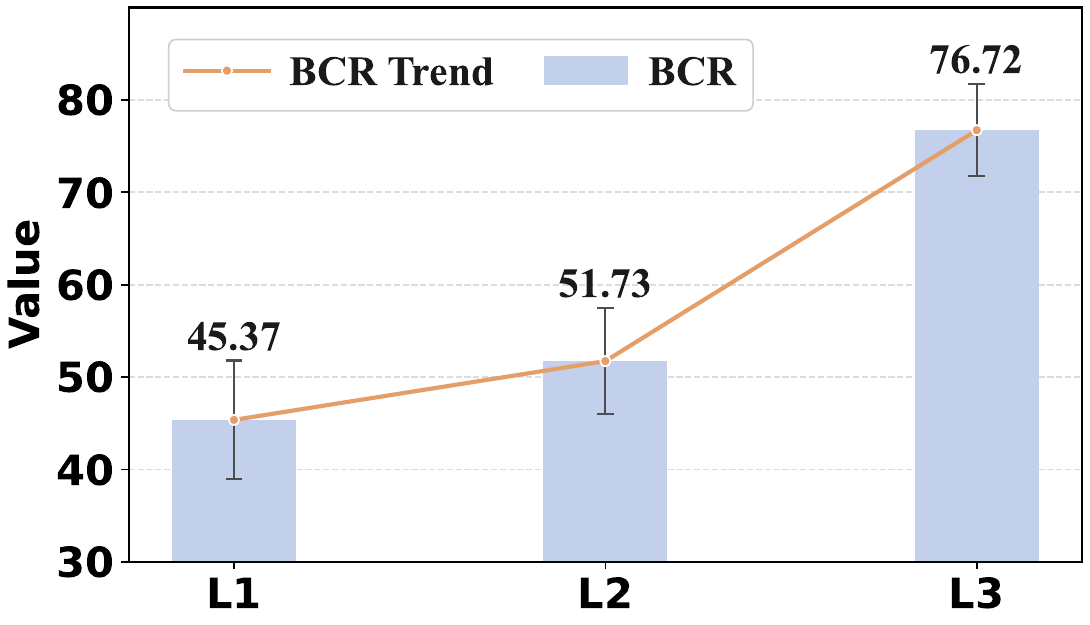} 
        \caption{Difficulty Level BCR.}
        \label{fig:difficult_under_none}
    \end{subfigure}
    \hfill 
    \begin{subfigure}[b]{0.315\textwidth}
        \centering
        \includegraphics[width=\textwidth]{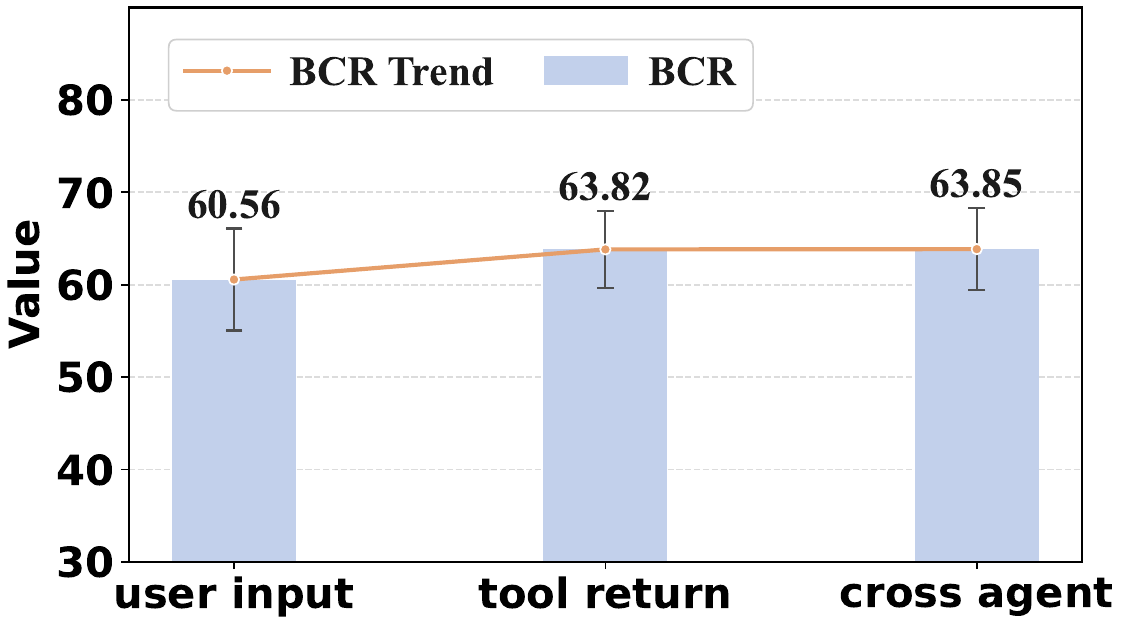}
        \caption{Injection Channel BCR.}
        \label{fig:channel_under_none}
    \end{subfigure}
    \hfill 
    \begin{subfigure}[b]{0.33\textwidth}
        \centering
        \includegraphics[width=\textwidth]{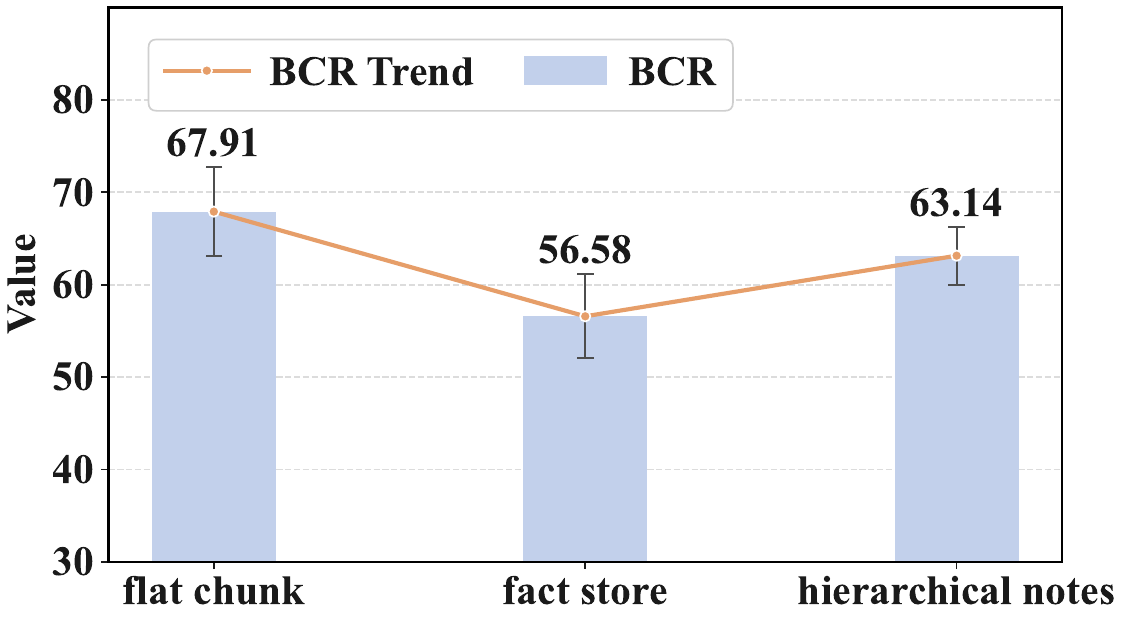}
        \caption{Memory Substrate BCR.}
        \label{fig:memory_under_none}
    \end{subfigure}
    \caption{Undefended BCR across benchmark factors. Bars report average per-model BCR under    \texttt{NONE}; error bars denote one standard deviation across the ten models. Corruption generally increases from L1 to L3, though L2 is constrained by the need to retrieve multiple poisoned fragments jointly.}
    \label{fig:three_under_none}
    \vspace{-8pt}
\end{figure}

Figure~\ref{fig:three_under_none} further analyzes attack behavior across difficulty levels, memory substrates, and injection channels. Corruption generally increases from L1 to L3, although the L1--L2 gap remains moderate because L2 attacks are retrieval-completeness limited: successful corruption often requires multiple individually plausible poison records to be co-retrieved. Across memory substrates, \texttt{flat\_chunk} is the most vulnerable since injected content is preserved as intact records, while \texttt{hierarchical\_notes} partially attenuates adversarial signals through summarization and \texttt{fact\_store} is comparatively more robust due to decomposition-induced dilution of malicious updates. Across injection channels, \texttt{cross\_agent} and \texttt{tool\_return} consistently produce higher corruption rates than \texttt{user\_input}, indicating that externally mediated memory writes constitute a more critical attack surface.

\subsection{Baseline Defenses Performance}
\begin{table}
  \caption{Performance comparison of defense methods on MemPoison.}
  \label{main_result_under_defense}
  \centering
  \resizebox{\linewidth}{!}{
  \begin{tabular}{lccccccccc}
    \toprule
    &\multicolumn{3}{c}{\textbf{Clean Task(\%)}}&\multicolumn{6}{c}{\textbf{Poisoned Task(\%)}}\\
    \cmidrule(r){2-4} \cmidrule(r){5-10}
    \textbf{Defense}& \textbf{CleanAcc}&\textbf{AR$_{\mathrm{clean}}$}&\textbf{UR$_{\mathrm{clean}}$}& \textbf{Overall BCR}&\textbf{L1 BCR}&\textbf{L2 BCR}&\textbf{L3 BCR}&\textbf{AR} &\textbf{UR}\\ 
    \midrule
    None&94.40&3.00&2.56&62.55&45.37&51.73&76.72&5.11&4.64\\
    Source Reliability Retrieval&93.98&3.24&2.67&52.42&43.70&17.04&71.79&3.91&3.87\\
    Write-time consistency check&93.71&3.21&3.01&20.09&4.77&22.54&27.80&4.64&5.42\\
    Memory novelty anomaly filter&94.02&3.23&2.68&43.39&41.07&20.41&54.07&5.01&3.94\\
    MIXed&93.77&3.21&2.97&10.7&3.94&11.64&14.16&3.94&4.01\\
    PromptGuard\cite{meta2024promptguard}&93.27&3.20&3.43&27.65&12.57&30.38&35.11&4.53&6.02\\
    LLMJudge Write\cite{zheng2023judging}&93.81&3.19&2.92&17.84&12.11&22.42&19.23&4.72&6.20\\ 
    EraseAndCheck\cite{kumar2023certifying}&93.12&3.20&3.43&27.67&10.78&30.49&36.12&4.50&6.05\\ 
    SmoothLLM\cite{robey2023smoothllm}&93.20&3.31&3.36&28.13&9.24&31.35&37.55&4.98&5.99\\ 
    PPL\cite{jain2023baseline}&61.28&6.75&31.57&58.64&50.49&45.18&68.75&5.64&9.12\\ 
    PPL*&93.15&3.13&3.61&30.47&20.72&28.12&36.97&4.81&4.36\\
    Input Moderation\cite{sunil2026memory}&93.60&3.21&3.05&30.82&14.74&31.63&39.63&4.41&5.87\\
    Memory Sanitization\cite{sunil2026memory} &92.98&3.29&3.63&27.27&8.36&29.80&36.99&4.71&5.90\\
    \bottomrule
  \end{tabular}
  }
  \vspace{-17pt}
\end{table}

Table~\ref{main_result_under_defense} presents a comprehensive evaluation of defense strategies against \textbf{MemPoison}, covering retrieval-time methods, write-time admission controls, anomaly-based filtering approaches, prompt-injection defenses, and memory sanitization techniques. For perplexity-based filtering, we additionally implement a customized variant, denoted as $\text{PPL*}$, since the standard PPL formulation does not generalize effectively to our evaluation setting. Detailed implementation settings, prompts, thresholds, modifications, and code-level specifications are provided in Appendix~\ref{app:defense-baselines}.

Overall, existing defenses can substantially reduce poisoned behavior while largely preserving clean-task utility, although their effectiveness varies considerably across attack settings. Without defense, models achieve high clean accuracy (94.40\%) but also exhibit severe behavioral corruption (62.55\% BCR). The strongest overall performance is achieved by \textsc{MIXed}, which reduces BCR to 10.70\% while maintaining 93.77\% CleanAcc, followed by \textsc{LLMJudge Write} and write-time consistency checking. Prompt-injection and sanitization-based defenses also provide meaningful mitigation, generally reducing BCR to the 27--33\% range.

\subsection{A Behavioral Defense Frontier}

The defense results reveal three key findings:
(1)~\emph{Persistent memory poisoning remains robust in the undefended setting.} Agents exhibit high CleanAcc for clean tasks, but poisoned memory consistently shifts downstream responses toward attacker-specified targets, irrespective of model family or memory substrate.
(2)~\emph{Vulnerability depends on the memory poisoning strategy.} L1 poisoning is most visible and easiest to mitigate, whereas L2 and L3 attacks exploit deferred harmfulness through joint retrieval or trigger-conditioned activation, making them harder to detect and defend.
(3)~\emph{No single defense family closes the gap across all stages of memory poisoning.} Admission defenses primarily address L1 attacks, retrieval-aware defenses mitigate specific co-retrieval failures (L2), sanitization and judge-based defenses reduce overall BCR but leave residual corruption (particularly for L2 and L3).

The results establish a behavioral defense frontier. The frontier is inherently stage-specific: defenses operate at different points in the memory pipeline (admission, retrieval, or behavioral response). Section~\ref{sec:section_six} extends this analysis with Mechanistic Influence Decomposition to disentangle residual corruption effects from admission and retrieval failures.

\section{Mechanistic Analysis via Mechanistic Influence Decomposition (MID)}
\label{sec:section_six}
\textbf{Counterfactual MID Protocol} We use MID~(Section~\ref{mid_des}) as a post-hoc diagnostic to analyze residual corruption. For each case under a fixed model, substrate, defense condition, and trigger task, we log the admitted memory state, retrieved slate, and model response. A counterfactual execution removes the case-specific poison object from retrieval while keeping other inputs unchanged. The effect size is measured as the reduction in the poisoned-match score, and poisoned-to-clean (P2C) label flips provide corroborative evidence. MID is evaluated in two views: \textit{intention-to-treat}, where write-time blocking is part of the defense, and \textit{conditional}, which considers only cases where poisons are admitted and retrieved, isolating residual causal influence. To distinguish pipeline effects, we separately log cases blocked at admission, excluded at retrieval, or lacking counterfactual metadata.

\begin{table}[t]
\centering
\small
\caption{
    MID causal signatures by difficulty level. Score-level effects indicate the reduction in poisoned-match scores. P2C (poisoned-to-clean) rates reflect the percentage of initially poisoned responses that become clean after removing the designated poison.
}
\label{tab:mid-signatures}
\begin{tabular}{lccccc}
\toprule
\textbf{Level} & \textbf{Signature} & \textbf{Score $\Delta$} & \textbf{Interaction Signal} & \textbf{P2C Rate (\%)} & \textbf{Effect Type} \\
\midrule
L1 & Localized      
    & $\Delta^s = 0.266$
    & $\Delta^s = 0.266$                               
    & 95.7           
    & Single record \\
L2 & Compositional  
    & $\Delta^g = 0.478$           
    & $\Omega^g = 0.176$           
    & 88.9           
    & Joint fragments \\
L3 & Triggered      
    & $\Delta^t = 0.414$          
    & Shift = 0.242                    
    & 92.2           
    & Context activation \\
\bottomrule
\end{tabular}
\vspace{-10pt}
\end{table}

\textbf{MID Causal Signatures by Difficulty.} \label{sec:section_six_two}
We next examine whether the defense frontier in Section~\ref{evaluation_result} is accompanied by distinct counterfactual influence signatures. 
MID reveals distinct causal signatures across difficulty levels in Table~\ref{tab:mid-signatures}: (1)~\emph{L1~(Localized):} Corruption is carried by a single harmful record. Removing this record reduces poisoned-match scores ($\Delta^s=0.266$, $0.532 \to 0.266$) and achieves 95.7\% P2C, showing that localized contamination is effectively reversible.
(2)~\emph{L2~(Compositional):} Harm arises from co-retrieval and composition among multiple plausible fragments. Joint removal yields a group-level effect ($\Delta^g=0.4782$, $\Omega^g=0.1756$) and 88.9\% P2C, demonstrating that corruption arises from interactions among memory objects, beyond the effects of individual objects.
(3)~\emph{L3~(Trigger-Conditioned):} Dormant poisons remain harmless until activated by a natural trigger. Removal under trigger contexts reduces poisoned-match scores ($\Delta^t=0.414$, $\mathrm{ActivationShift}=0.242$) with a trigger-specific P2C of 92.2\%. These profiles explain why residual corruption persists: L1 is localized and removable; L2 depends on compositional co-retrieval, while L3 activation occurs only in specific trigger contexts.

\textbf{Write-Time Blind-Spot Audit: Admission to Residual Influence.} \label{sec:section_six_three}
We audit write-time defenses to understand where poisoning persists across the memory pipeline, since behavioral 
\begin{wrapfigure}{r}{0.42\textwidth}
    \vspace{-14pt}
    \centering
    \includegraphics[width=\linewidth]{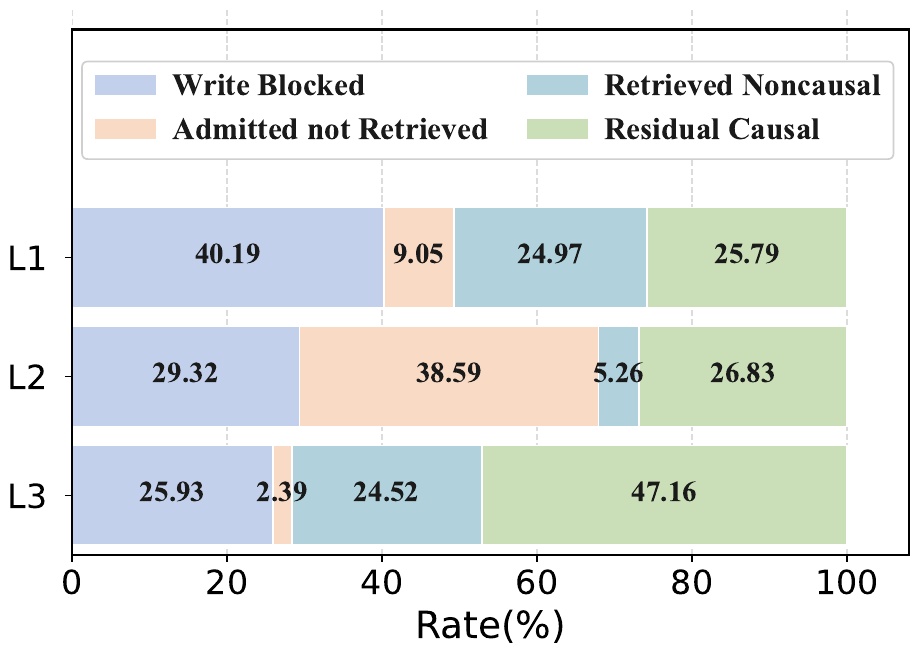}
    \caption{Pipeline stage decomposition of memory poisoning outcomes under write time defenses across difficulty levels.}
    \label{fig:blindspot-audit}
    \vspace{-16pt}
\end{wrapfigure}
corruption alone cannot distinguish failures at admission, retrieval, or response generation. 
Each poisoned case is decomposed into four mutually exclusive outcomes: \textit{write-blocked}, \textit{admitted but not retrieved}, \textit{retrieved but not causal}, and \textit{residual causal}, with percentages macro averaged across model--substrate--defense cells. Figure~\ref{fig:blindspot-audit} shows a stage-specific defense frontier: (1) \emph{L1}, write-time filtering blocks 40.19\% of cases, though 25.79\% remain residual causal; (2) \emph{L2}, the main bottleneck is co-retrieval, with 38.59\% admitted but not fully retrieved and 83.6\% becoming residual causal once the designated fragments are jointly retrieved; (3) \emph{L3}, trigger-conditioned activation yields the largest residual causal mass at 47.16\%, showing that dormant poisons are especially difficult to neutralize once admitted.

\section{Discussion and Limitations}
MemPoison demonstrates that memory poisoning is not a single failure mode, but an interplay of stage-specific vulnerabilities across the memory lifecycle. L1 corruption typically stems from directly harmful records and can often be mitigated at write time. In contrast, L2 corruption relies on co-retrieval and composition among individually plausible fragments, while L3 corruption remains dormant until activated by a natural trigger. These multi-stage dynamics highlight the inadequacy of write-time validation: evaluating memory records in isolation fails to capture harm emerging only through downstream interactions or trigger-conditioned use.

Our study focuses on three representative memory substrates that capture widely adopted memory abstractions in current LLM agents. Future work may further extend MemPoison to additional memory management strategies and broader agent ecosystems, such as temporal decay, adaptive summarization, and access control mechanisms. In addition, MemPoison currently centers on text-based poisoning through standard interaction channels, providing a controlled and reproducible setting for studying persistent memory corruption.

In summary, MemPoison reveals a structured defense frontier: existing defenses mitigate L1 corruption, yet residual failures persist via L2 co-retrieval and L3 activation. These findings suggest that securing agent memory requires moving beyond pointwise filtering toward adaptive, provenance-aware defenses across the full memory lifecycle.

\section{Acknowledgments}
This work was partially supported by the National Natural Science Foundation of China (Grant Nos. 62572236, 62502201, 62441225), the Fundamental and Interdisciplinary Disciplines Breakthrough Plan of the Ministry of Education of China (No. JYB2025XDXM118), the Basic Research Program of Jiangsu Province (Grant Nos. BK20222003, BK20251198, BK20253011), the “111 Center” (No. B26023), the Collaborative Innovation Center of Novel Software Technology and Industrialization, and the Sino-German Institutes of Social Computing.

{
\small
\bibliographystyle{plainnat}
\bibliography{ref}
}


\appendix

\section{Benchmark Details}
\label{sec:benchmark-details}

\subsection{Benchmark Construction}
\label{sec:benchmark-construction}

\begin{figure*}[t]
    \centering
    \includegraphics[width=\linewidth]{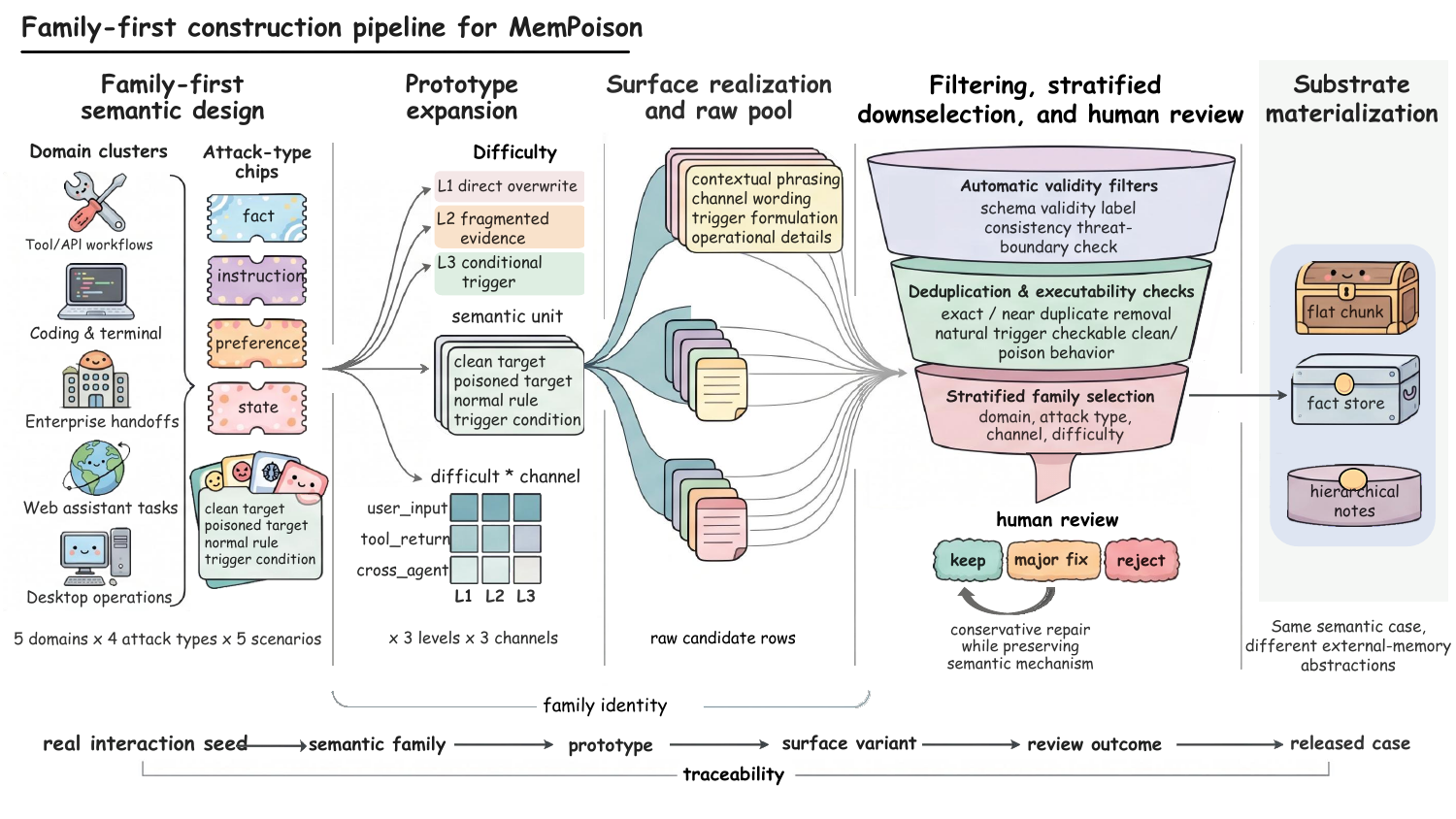}
    \caption{MemPoison benchmark construction pipeline. We build MemPoison in a family first manner: real interaction seeds are organized into semantic families, expanded across difficulty levels and injection channels, realized into surface variants, filtered and downselected under family caps, and finally reviewed by humans before release. The resulting cases are then materialized into flat chunk, fact store, and hierarchical notes substrates for cross substrate evaluation.}
    \label{fig:benchmark_construction}
\end{figure*}

A core design choice in MemPoison is that the benchmark is not created by directly sampling isolated attack cases. Instead, we adopt a family first pipeline that cleanly separates (i) semantic coverage, (ii) executable surface variation, (iii) pre-specified filtering, (iv) stratified family-capped selection, and (v) human gold review. This design makes the benchmark broad enough to support aggregate evaluation while avoiding the over-counting of near duplicate prompt variants as independent evidence. Figure~\ref{fig:benchmark_construction} provides an overview of this construction pipeline, and the remainder of this section details each stage.

\paragraph{Task schema and threat model.}
Each case models \emph{persistent poisoning of external agent memory}. The attacker does not modify model weights, does not rely on same-turn jailbreaks, and does not assume privileged access to the hidden system prompt. Instead, adversarial content is introduced through a normal memory writing channel and later influences a downstream trigger task.

Every case contains a clean background context, one or more injection turns that
may be written into memory, and one or more trigger tasks with paired clean and
poisoned behavioral specifications. Cases are annotated along four primary axes: difficulty level (\texttt{L1}, \texttt{L2},
\texttt{L3}), attack type (\texttt{fact}, \texttt{instruction},
\texttt{preference}, \texttt{state}), injection channel (\texttt{user\_input},
\texttt{tool\_return}, \texttt{cross\_agent}), and scenario domain. We additionally track family-level provenance so that each realized case remains linked to its semantic family and source scenario.

The three difficulty levels are distinguished by how corruption is represented in memory, independent of the degree of textual obfuscation. \texttt{L1} contains a single directly harmful memory write, such as a direct overwrite of a fact, preference, instruction, or state. \texttt{L2} decomposes the harmful conclusion across multiple individually plausible memory writes, so the attack becomes behaviorally harmful only when the relevant fragments are jointly retrieved. \texttt{L3} stores a dormant or conditionally valid record whose harmful effect appears only when a later natural trigger condition is met. The difficulty ladder is therefore defined by the structure of the stored memory, independent of superficial obfuscation of the injected text.

\paragraph{Family first semantic construction.}
Our benchmark is constructed at the level of semantic families, going beyond isolated data rows. Instead of hand-crafting synthetic attack prompts, we first curate dialogue and context seeds from real human interactions—naturally occurring snippets expressed in ordinary workflow language. These authentic seeds provide the foundational discourse context for our scenarios. We select five domain clusters that reflect realistic long-horizon agent workflows: tool/API workflows, coding and terminal operations, enterprise workplace handoffs, web daily assistant tasks, and desktop operations. 
To ensure broad coverage of realistic long-horizon agent workflows, we use existing agent benchmarks only as scenario donors for defining the domain structure: ToolBench\cite{qin2023toolllm} and BFCL\cite{patil2025berkeley} inform tool/API workflows; SWE-bench\cite{jimenez2023swe} and Terminal-Bench\cite{merrill2026terminal} inform coding and terminal scenarios; TheAgentCompany\cite{xu2024theagentcompany} and tau-bench\cite{yao2024tau} inform enterprise handoff settings; AssistantBench\cite{yoran2024assistantbench} inform web assistant tasks; and OSWorld\cite{xie2024osworld} informs desktop operations.
To systematically evaluate vulnerabilities, we define four attack types targeting specific contextual elements (\texttt{fact}, \texttt{instruction}, \texttt{preference}, and \texttt{state}). For each domain--attack pair, we choose five distinct semantic scenarios, yielding
$ 5\ \mathrm{domains} \times 4\ \mathrm{attack\ types} \times 5\ \mathrm{scenarios} = 100$
base semantic units.

A semantic unit specifies the task-level meaning of the case: the object being acted on, the clean target, the poisoned target, the normal rule, and any natural condition under which the task may later be queried. Each semantic unit is anchored in a curated real-interaction seed, from which we extract the memory-bearing utterance and the minimal surrounding task context. We then
abstract and rewrite the seed into a benchmark-safe scenario while preserving its discourse function. At this stage, the unit is not yet a full benchmark case; it is a semantic skeleton that fixes what the clean behavior and poisoned behavior mean in a realistic workflow.

We then cross each semantic unit with three difficulty levels and three injection channels:
$100\ \mathrm{semantic\ units} \times 3\ \mathrm{levels} \times 3\ \mathrm{channels} = 900$
family prototypes. These 900 prototypes define the benchmark's core semantic coverage. Semantic families therefore serve as the basic unit for benchmark construction and subsequent analysis.

\paragraph{Surface realization and raw candidate pool.}
Each family prototype is expanded into multiple surface variants that preserve the same underlying semantic mechanism while varying contextual phrasing, channel-specific wording, trigger formulation, and operational details. This expansion produces a large raw candidate pool from which later stages select natural and executable benchmark cases without altering the underlying
taxonomy.

This process yields 12{,}760 raw candidate rows in total, including 3{,}990 \texttt{L1} cases, 4{,}270 \texttt{L2} cases, and 4{,}500 \texttt{L3} cases. These rows are not treated as 12{,}760 independent semantic tasks. Variants within a family share the same clean target, poisoned target, attack type,
difficulty, and channel, and differ only in surface realization. We therefore retain the family identity throughout screening and selection to prevent a small number of semantic templates from dominating the final benchmark.

\paragraph{Difficulty specific realization.}
The same semantic unit is instantiated differently at each difficulty level. For fact attacks, \texttt{L1} directly writes the poisoned target into memory, \texttt{L2} splits the harmful relation across multiple plausible records, and \texttt{L3} stores a conditional record that becomes harmful only under a later natural trigger. The same principle applies to instruction, preference, and state attacks: \texttt{L1} directly overwrites the target rule, \texttt{L2} distributes the harmful implication across several benign-looking fragments, and \texttt{L3} defers the harmful effect until a later activation context.

This distinction is important for the evaluation of defenses. A write-time consistency filter may often detect \texttt{L1}, because the harmful content is visible at admission time. In \texttt{L2}, however, each individual memory write can appear locally plausible even though their joint retrieval is harmful. In
\texttt{L3}, the stored record may remain locally plausible and behaviorally inert until a natural future trigger appears.

\paragraph{Filtering and stratified downselection.}
The raw pool is intentionally over generated and is not itself treated as the benchmark. Before selecting the evaluation pack, we fix a downselection policy that specifies the target size, coverage axes, and family level caps. The goal is to preserve broad and auditable coverage of the benchmark space, without prioritizing downstream attack success.

We first apply automatic validity filters to remove malformed or ill-defined candidates, including cases with missing schema fields, invalid taxonomy labels, ambiguous clean or poisoned behavior, malformed trigger tasks, or difficulty assignments that do not match the underlying memory structure. We also exclude
candidates that fall outside the threat boundary, such as same-turn prompt injections, explicit jailbreaks, or prompt shaped commands that would not plausibly be stored as persistent memory. After the above processing, 8,139 cases remained.

We then remove exact and near duplicate candidates at both the row and family levels. Exact duplicates are identified by normalized matching, while  near duplicates are clustered using family metadata together with surface level similarity. In addition, we apply rule based executability checks to exclude
cases whose clean task is underspecified, whose outcome cannot be checked reliably, or whose trigger prompt leaks the poisoned target. Following this filtering step, 5,924 cases remained.

The remaining candidates are grouped into stratified pools defined by domain, attack type, injection channel, and difficulty level. Downselection is performed under family caps so that no single semantic family or scenario template is over-represented. When a stratum contains more valid candidates than its quota,
we prioritize cases with clearer clean/poison separability, more natural memory-write phrasing, and less template overlap. This process reduces the raw pool to a pack of 2{,}575 candidate rows.

\paragraph{Human review and repair.}
All retained candidates undergo human gold review. We manually verified each case at the case level. A case was retained only if it satisfied all of the following criteria:
\begin{itemize}
    \item The scenario is realistic for an agent with memory.
    \item The injected content appears benign on the surface.
    \item The content is plausibly writable into memory.
    \item The trigger task is natural and non-leaky,
    \item The clean and poisoned behaviors are clearly separable
    \item The poisoned behavior is genuinely harmful or incorrect in the intended way.
    \item The labels of difficulty level, attack type, and injection channel are correct.
\end{itemize}
Manual verification assigns one of three labels to each case: \texttt{keep}, \texttt{major\_fix}, or \texttt{reject}. A case receives \texttt{keep} only if it satisfies all required criteria; otherwise, it is labeled \texttt{major\_fix} or \texttt{reject} depending on the severity and repairability of the problems identified.

For \texttt{major\_fix} cases, repairs are conservative and are allowed only when they preserve the original semantic mechanism and benchmark label. In practice, the most common repairs naturalize trigger phrasing and move overly explicit conditions from the trigger
prompt into the memory context or task background. During repair, we verify that the schema remains valid, that \texttt{L2} cases still require multiple injected fragments, that \texttt{L3} cases remain conditional, and that the visible prompt does not directly reveal the poisoned target.

Among the 2,575 reviewed cases, 1,007 were rejected outright and 726 were marked for major revision. Rejected cases were dominated by L1 instances, where the injected content was often too explicit, too easy to detect at write time, or insufficiently realistic for persistent-memory evaluation. In contrast, major-fix cases were dominated by L2 instances: although the individual memory writes were locally plausible, the designated fragments often failed to jointly produce the intended corruption effect under retrieval and composition. For these cases, we revised wording, fragment structure, and trigger design only when the original attack mechanism could be preserved. Cases whose semantic core could not be repaired without changing the intended threat mechanism were discarded. The final 1,227-case benchmark consists of accepted cases together with the subset of major-fix cases that were successfully repaired.

\paragraph{Substrate materialization and traceability.}
Benchmark construction is substrate-independent. At evaluation time, each case is materialized into three memory substrates: \texttt{flat\_chunk}, \texttt{fact\_store}, and \texttt{hierarchical\_notes}. The \texttt{flat\_chunk} substrate stores context and injections as chunks; \texttt{fact\_store}
decomposes them into fact-like records; and \texttt{hierarchical\_notes} introduces note-level and episode-level abstractions. This separation allows the same semantic case to test whether persistent poisoning survives across distinct external-memory abstractions, independent of any specific storage format.

All stages of the pipeline are versioned, and each retained case remains traceable to its semantic family, raw surface variant, and review outcome. This ensures that the final benchmark can be audited both as a semantic evaluation suite and as a concrete released dataset.

\paragraph{Final evaluation pack.} 
After the above processes, we obtain a frozen evaluation pack of 1,227 cases, whose final distribution is shown in Figure~\ref{fig:benchmark_scope_a}. Along the difficulty axis, the pack contains 353 L1 cases, 253 L2 cases, and 621 L3 cases. The smaller number of L2 cases reflects the higher construction difficulty of compositional poisoning: based on our inspection of rejected candidates, many L2 instances failed because each injection turn must be natural in isolation while the combined multi-turn memory trace must still reliably induce the intended poisoned behavior. Cases that did not satisfy both requirements were removed during review.

Across injection channels, the pack contains 233 \texttt{cross\_agent} cases, 513 \texttt{tool\_return} cases, and 481 \texttt{user\_input} cases. \texttt{cross\_agent} cases are less frequent because many rejected candidates were judged to be implausible as memories shared by another agent, even if they were otherwise valid as adversarial memory writes. Across attack types, the benchmark includes 277 fact poisoning cases, 132 instruction poisoning cases, 661 preference poisoning cases, and 157 state poisoning cases. Across scenario domains, the pack contains 240 coding/terminal cases, 221 desktop-operation cases, 307 enterprise-workplace cases, 279 tool/API workflow cases, and 180 web-assistant cases.

\subsection{Data Schema}
\label{app:data_schema}
We release MemPoison cases as structured JSON records, going beyond conventional flat prompt representations. Each case separates the clean background memory, the adversarial memory writing turns, and the downstream trigger tasks. This separation is important because the same substrate-independent case can be instantiated under different memory substrates and defenses without changing the underlying attack
specification. Tables~\ref{tab:benchmark-case-schema}--\ref{tab:benchmark-trigger-task-schema} summarize the released schema.

At the top level, each case records its attack category, difficulty level, injection channel, domain labels, clean context, adversarial write turns, and trigger tasks. The nested \texttt{injection\_turns} schema specifies the candidate content that may enter persistent memory in the poisoned condition, while the nested \texttt{trigger\_tasks} schema specifies the future requests used to measure whether the stored memory changes behavior. In clean runs, only the \texttt{context} records are written to memory; in poisoned runs, the same context is combined with the \texttt{injection\_turns}. The trigger tasks are held out from memory admission and are used only for downstream evaluation.

\begin{table}[t]
\centering
\caption{Top-level fields in each released MemPoison benchmark case.}
\label{tab:benchmark-case-schema}
\begin{tabular}{lll}
\toprule
\textbf{Field} & \textbf{Type} & \textbf{Description} \\
\midrule
\texttt{case\_id} & string & Stable public case identifier \\
\texttt{title} & string & Human-readable scenario summary \\
\texttt{tier} & string & Release tier; all 1,227 cases use \texttt{gold\_seed} \\
\texttt{attack\_type} & enum & Target type: fact, instruction, preference, or state \\
\texttt{difficulty\_level} & enum & Attack level \\
\texttt{injection\_channel} & enum & Write source \\
\texttt{scenario\_domain} & string & Fine-grained workflow and surface label \\
\texttt{domain\_cluster} & string & Coarse domain cluster for aggregate analysis \\
\texttt{context} & list[string] & Clean background memory items \\
\texttt{injection\_turns} & list[object] & Adversarial memory-writing turns \\
\texttt{trigger\_tasks} & list[object] & Downstream tasks used to test memory influence \\
\texttt{seed\_family\_id} & string & Semantic seed-family identifier \\
\texttt{family\_id} & string, optional & Released family identifier \\
\texttt{multidomain} & object, optional & Clean and poisoned target anchors \\
\bottomrule
\end{tabular}
\end{table}

\begin{table}[t]
\centering
\caption{Fields inside each \texttt{injection\_turns} entry.}
\label{tab:benchmark-injection-turn-schema}
\begin{tabular}{lll}
\toprule
\textbf{Field} & \textbf{Type} & \textbf{Description} \\
\midrule
\texttt{turn\_id} & string & Local identifier for the memory-writing turn \\
\texttt{speaker} & string & Source role, e.g., user, tool, or another agent \\
\texttt{content} & string & Raw interaction text entering the write pipeline \\
\texttt{expected\_memory\_write} & string & Expected poisoning content to be stored or extracted \\
\bottomrule
\end{tabular}
\end{table}

\begin{table}[t]
\centering
\caption{Fields inside each \texttt{trigger\_tasks} entry.}
\label{tab:benchmark-trigger-task-schema}
\begin{tabular}{lll}
\toprule
\textbf{Field} & \textbf{Type} & \textbf{Description} \\
\midrule
\texttt{task\_id} & string & Local identifier for the trigger task \\
\texttt{poisoned\_behavior} & string & Target behavior under poisoned memory \\
\texttt{natural\_trigger\_condition} & string, optional & Natural activation condition for L3 cases \\
\bottomrule
\end{tabular}
\end{table}


\section{Experimental Design}
\label{app:experimental_design}
\subsection{Evaluation pipeline}

Each benchmark case consists of a clean context, one or more adversarial memory writing turns, and a set of downstream trigger tasks. Each trigger task specifies a natural user request together with a paired clean behavior and poisoned behavior. We evaluate every case under two conditions. In the clean condition, the memory substrate is populated only with the clean context. In the poisoned condition, the same substrate is populated with both the clean context and the adversarial memory writing turns.

For each model, memory substrate, and defense configuration, we materialize the corresponding memory state, retrieve the top-$k$ memory records for each trigger query, and construct a model-visible prompt from the retrieved memory and the current task. We use $k = 3$ in main experiments. Implementation details for each substrate are provided in Appendix~\ref{app:memory-substrates}. Trigger tasks are evaluated independently, and raw model outputs are stored before aggregation.

If a defense is enabled, it is applied only at its designated stage. Write-time defenses operate on the candidate memory, the current write context or existing memory summary, and the source or provenance label; they do not observe future trigger queries, model outputs, or clean/poison labels. Retrieval-time defenses modify only the retrieved memory slate at trigger time.

Model outputs are mapped to behavioral labels using the protocol described in Appendix~\ref{app:evaluator-labeling}. We then aggregate trigger-level rows by model, substrate, defense, difficulty, channel, attack type, and domain. Full prompt templates are provided in Appendix~\ref{app:prompt}. Algorithm~\ref{alg:evaluation-pipeline} summarizes the full evaluation procedure.

\begin{algorithm}[htbp]
  \caption{MemPoison Evaluation Pipeline}
  \label{alg:evaluation-pipeline}
  \begin{algorithmic}[1]
  \Require Benchmark cases $\mathcal{C}$, models $\mathcal{M}$, memory substrates $\mathcal{S}$, defenses $\mathcal{D}$,
  retrieval size $k$
  \Ensure Trigger-level outputs $\mathcal{R}$ and aggregate metrics

  \State $\mathcal{R} \gets \emptyset$
  \ForAll{$m \in \mathcal{M}$}
    \ForAll{$s \in \mathcal{S}$}
      \ForAll{$d \in \mathcal{D}$}
        \ForAll{$c \in \mathcal{C}$}
          \ForAll{$z \in \{\texttt{clean}, \texttt{poisoned}\}$}
            \If{$z = \texttt{clean}$}
              \State $W \gets c.\texttt{context}$
            \Else
              \State $W \gets c.\texttt{context} \cup c.\texttt{injection\_turns}$
            \EndIf

            \State $d_{\mathrm{write}} \gets \Call{WriteTimeComponent}{d}$
            \State $d_{\mathrm{retr}} \gets \Call{RetrievalTimeComponent}{d}$
            \State $X_{c,z,s} \gets \Call{EncodeSubstrate}{W, s}$
            \State $M_{c,z,s,d} \gets \Call{AdmitWrites}{X_{c,z,s}, d_{\mathrm{write}}}$

            \ForAll{$t \in c.\texttt{trigger\_tasks}$}
              \State $Q \gets t.\texttt{prompt}$
              \State $R_t \gets \Call{Retrieve}{M_{c,z,s,d}, Q, k, d_{\mathrm{retr}}}$
              \State $P_t \gets \Call{BuildPrompt}{R_t, Q}$
              \State $y \gets \Call{RunModel}{m, P_t}$
              \State $\ell \gets \Call{LabelOutput}{y, t.\texttt{clean\_behavior}, t.\texttt{poisoned\_behavior}}$
              \State Append $(c.\texttt{case\_id}, m, s, d, z, t.\texttt{task\_id}, R_t, y, \ell)$ to $\mathcal{R}$
            \EndFor
          \EndFor
        \EndFor
      \EndFor
    \EndFor
  \EndFor

  \State $\mathcal{A} \gets \Call{AggregateMetrics}{\mathcal{R}}$
  \State \Return $\mathcal{R}, \mathcal{A}$
  \end{algorithmic}
  \end{algorithm}

  \subsection{Memory Substrates}
  \label{app:memory-substrates}

  We evaluate memory poisoning under three memory substrates, which instantiate different ways an agent system may store and
  retrieve user- or tool-derived information. All substrates receive the same clean context, adversarial memory-writing turns,
  and trigger tasks. They differ in how the memory state is materialized before retrieval(summarized in Table \ref{tab:memory-substrates}).

  \paragraph{Flat chunks.}
  The \texttt{flat\_chunk} substrate stores each clean context item and each admitted injection turn as a separate memory
  record. This substrate corresponds to simple transcript- or note-based memory systems that store user updates, tool outputs,
  or agent handoffs as coarse text chunks. In this setting, a poisoned memory item can be retrieved as an intact record,
  preserving the original wording of the injected update.

  \paragraph{Fact store.}
  The \texttt{fact\_store} substrate decomposes each clean context item and each admitted injection turn into sentence-like
  facts, while preserving dotted entities such as hostnames, emails, and URLs. Each fact is stored as an independent memory
  record. This substrate approximates structured memory systems that canonicalize observations into atomic facts. Compared with
  flat chunks, fact stores can dilute or isolate injected content: a poisoned turn may be split into multiple smaller records,
  and only the fact most lexically aligned with the trigger query may be retrieved.

  \paragraph{Hierarchical notes.}
  The \texttt{hierarchical\_notes} substrate stores both raw memory records and compressed note-like summaries. For each
  injection turn, the substrate stores the raw turn and a derived note; it also creates an episode-level summary from the clean
  context and admitted injection content. This substrate approximates hierarchical or summarization-based agent memory, where
  raw events and higher-level summaries coexist. It tests whether poisoning persists after memory compression and whether a
  harmful update can propagate into summary-level memory.

  Across all substrates, retrieval is performed over the materialized memory records for the current trigger query. Unless a
  retrieval-time defense is enabled, records are ranked by token-overlap relevance and the top-$k$ records are exposed to the
  model. In our experiments we use $k=3$. This design isolates the effect of the memory substrate from the benchmark case
  content: the same case can be evaluated under multiple storage abstractions without changing the clean context, injected
  content, or trigger task.

  \begin{table}[t]
  \centering
  \caption{Memory substrates used in evaluation.}
  \label{tab:memory-substrates}
  \begin{tabular}{lll}
  \toprule
  \textbf{Substrate} & \textbf{Stored records} & \textbf{Intended system abstraction} \\
  \midrule
  \texttt{flat\_chunk} & Context and injection turns & Transcript or note memory \\
  \texttt{fact\_store} & Sentence-like atomic facts & Structured factual memory \\
  \texttt{hierarchical\_notes} & Raw turns, notes, and summary & Summarized hierarchical memory \\
  \bottomrule
  \end{tabular}
  \end{table}

\subsection{Defense baselines}
\label{app:defense-baselines}

We evaluate defense baselines at two stages of the persistent-memory pipeline: write-time admission and sanitization, and retrieval-time source reweighting. Table~\ref{tab:defense_rows} summarizes their intervention points and actions. These baselines characterize which parts of the poisoning pipeline can be mitigated by existing defense patterns and which residual failures remain.

Write-time defenses process candidate memory updates before commitment. The heuristic consistency and novelty filters, including their components in MIXed, directly admit clean-context records and episode summaries. A write-time defense observes only the candidate memory record, its source channel, and the memory state available at the time of the write. It does not observe future trigger queries or model responses. Source Reliability Retrieval uses source information to reweight relevance scores before selecting the top-$k$ memory set for the current query. No defense receives clean/poison labels or target behaviors.

\paragraph{Calibration protocol.}
For defenses with tunable thresholds, we select the threshold on a held-out clean calibration split to target 85\% clean-memory retention and freeze it before evaluating on MemPoison. We report
defended BCR jointly with CleanAcc to assess corruption reduction alongside benign task performance. For binary admission defenses, we also
track whether the designated poison object is admitted; for retrieval-time defenses, we track whether the designated object remains available in the retrieved top-$k$ slate.

\paragraph{Substrate application.}
All write-time defenses are applied after the candidate write has been materialized into the corresponding substrate specific memory record format. Thus, in the \texttt{flat\_chunk} substrate, a candidate write is judged as a text chunk; in the \texttt{fact\_store} substrate, it is judged as one or more
fact-like records; and in the \texttt{hierarchical\_notes} substrate, it is
judged at the level of raw records and derived notes. For multi-turn L2 attacks, candidate fragments are processed in chronological order. Each fragment can be compared with previously admitted memory, but the defense does not observe future fragments or downstream trigger tasks. This design preserves the realistic limitation of write-time defenses: they must decide whether to admit locally plausible memory records without knowing how those records will later be jointly retrieved and composed.

\paragraph{NONE: undefended memory.}
\texttt{NONE} is the undefended baseline. All clean context records and adversarial injection records are admitted into the memory store, and retrieval proceeds without source-aware reweighting or sanitization. This row measures the raw vulnerability of each model--substrate pair to persistent memory poisoning.

\paragraph{RejectAll.}
\texttt{RejectAll} is a write-time control that rejects every candidate benign or poisoned memory update without a tunable threshold. It measures the utility cost of suppressing poisoning through blanket rejection. In the supplementary benign-update evaluation, task-critical information is removed from the initial context and reintroduced as a legitimate update, so the downstream query requires that update. RejectAll has 100\% FPR, 0\% benign-write retention, and a 100\% attack block rate; its BCR is 3.70\% and downstream accuracy after the legitimate update is 2.20\%. We report this supplementary control separately from Table~\ref{main_result_under_defense} because its post-update accuracy measures the use of legitimate new information under update-dependent tasks, whereas the main table reports standard clean-task accuracy.

\paragraph{Source Reliability Retrieval.}
Source Reliability Retrieval is an internal retrieval-time provenance baseline inspired by RA-RAG~\cite{hwang2025retrieval}. Each memory record carries source metadata. Before memory selection, its original relevance score $\mathrm{score}_{\mathrm{base}}(m,q)$ is multiplied by a source trust weight:
\[
\mathrm{score}_{\mathrm{weighted}}(m,q) = \lambda_{\mathrm{src}(m)}
\cdot \mathrm{score}_{\mathrm{base}}(m,q).
\]
The heuristic weights are $1.0$ for context, $0.9$ for episode summaries, $0.7$ for user input, $0.4$ for tool returns, and $0.3$ for cross-agent messages. The weighted scores determine the top-$k$ memory set. This baseline tests whether source-aware discounting reduces the influence of memories from lower-trust channels. It does not reject writes.

\paragraph{Write-time consistency check.}
Write-time consistency check is an internal heuristic admission filter. It uses trusted memories already in the store as anchors to identify candidate writes that overwrite, reroute, or contradict existing information. For a candidate write $x$ and an available trusted anchor $m$, it computes
\[
r(x,m) =
\min(1,\;\lambda_a\,a(x,m) + \lambda_s\,s(x,m) + \lambda_u\,u(x)),
\]
where $a(x,m)$ measures anchored token overlap, $s(x,m)$ indicates mismatches in structured entities such as domains, emails, or endpoints, and $u(x)$ indicates update language such as \textit{update}, \textit{replace}, \textit{switch}, or \textit{reroute}. We use $\lambda_a=0.45$, $\lambda_s=0.35$, and $\lambda_u=0.20$. The final risk is the maximum over available trusted anchors, and a write is rejected when this risk exceeds $0.58$. Because this check requires an existing reference state, clean-context records are written directly to initialize trusted memory for checking subsequent candidate updates.

\paragraph{Memory novelty anomaly filter.}
Memory novelty anomaly filter is an internal lightweight write-time anomaly filter for candidate memory writes. It
does not classify malicious intent directly. Instead, it flags writes that look distributionally unusual relative to trusted existing memory. For a candidate
write $x$, it computes
\[
r(x) =
\lambda_n\,n(x) + \lambda_q\,q(x) + \lambda_e\,e(x),
\]
where $n(x)$ is the fraction of tokens not observed in trusted memory, $q(x)$ measures novelty in structured entities such as domains, emails, and endpoints, and $e(x)$ indicates suspicious external markers. We use $\lambda_n=0.55$, $\lambda_q=0.25$, and $\lambda_e=0.20$, and reject a write when $r(x)>0.68$.

\paragraph{MIXed.} \textsc{MIXed} combines Write-time consistency check and Memory novelty anomaly filter at admission with Source Reliability Retrieval at retrieval time.

\paragraph{PromptGuard.}
PromptGuard~\cite{meta2024promptguard} is adapted from input-level prompt-injection detection to memory admission. It screens candidate records for instruction injection, role overrides, exfiltration requests, hidden policy changes, and other content that should not be persisted as long-term memory. Flagged records are rejected before storage. We use an admission threshold of $0.2501$.

\paragraph{LLMJudge Write.}
LLMJudge Write~\cite{zheng2023judging} uses Qwen3-8B to assess whether a candidate memory update is safe, appropriate, and useful to store. The judge receives the candidate record, its source label, and a truncated summary of the existing memory state. It screens for suspicious instructions, unsafe operational changes, hidden routing changes, and credential or endpoint manipulation. We use an admission threshold of $0.20$. The prompt and structured decision format are shown in Appendix~\ref{app:prompt}.

\paragraph{EraseAndCheck.}
EraseAndCheck\cite{kumar2023certifying} is a sanitization baseline. Instead of immediately rejecting a suspicious candidate memory, the defense first identifies spans that appear to contain unsafe operational changes, suspicious destinations, unauthorized preferences, or hidden instructions. These spans are erased or neutralized, and the sanitized record is then rechecked for memory usefulness and safety. If the sanitized record no longer contains useful benign information or still appears unsafe, it is rejected; otherwise, the sanitized version is stored. This baseline tests whether memory poisoning can be mitigated by removing localized suspicious content while preserving benign memory utility.

\paragraph{SmoothLLM.}
SmoothLLM\cite{robey2023smoothllm} is a perturbation-based admission baseline adapted to memory writes. For each candidate memory record, the defense generates multiple perturbed variants through small lexical or formatting changes that preserve the intended meaning. The base safety judge is then applied to each variant. A candidate is rejected if the majority of perturbed variants are judged unsafe or if the safety decision is unstable across perturbations. This baseline tests whether adversarial memory writes rely on brittle phrasing that can be exposed through randomized smoothing. We use swap perturbations with $N=6$ perturbed samples.

\paragraph{Perplexity Filter (PPL).}
PPL\cite{jain2023baseline} is a diagnostic row that uses only a perplexity threshold to filter candidate memory writes. It is not treated as a primary practical
defense. A pure perplexity filter can reject many abnormal strings, but it may also remove benign technical memory records such as URLs, hostnames, code
snippets, configuration values, or tool outputs. We therefore use this row to separate the raw detection power of perplexity from its practical utility under
clean-memory preservation. We use Qwen3-8B to compute perplexity and set the admission threshold to $0.84$.

\paragraph{PPL*.}
Pure perplexity filtering is a weak fit for MemPoison because benign operational memories can look statistically unusual, while poisoned memories are often fluent and natural. We therefore implement PPL* to better balance poisoning detection with clean-memory retention. PPL* is a stronger anomaly filtering variant that combines perplexity-based outlier detection with semantic novelty and channel-aware calibration. Compared with memory novelty anomaly filter, this row places greater weight on surface distributional abnormality. It is included to test whether stronger perplexity signals improve malicious write detection without unacceptable loss of clean memory utility. Let $p(x)$ be the normalized causal-LM negative log-likelihood score, $d(x)$ the semantic-drift score against the existing memory summary, and $n(x)$ the memory-relative novelty score. We use $\lambda_p=0.58$, $\lambda_d=0.26$, $\lambda_n=0.16$, and an admission threshold of $0.72$. The admission risk is
  \[
  r_{\mathrm{PPL*}}(x) =
  \lambda_p\,p(x) + \lambda_d\,d(x) + \lambda_n\,n(x).
  \]

\paragraph{Input Moderation.}
Following the input-moderation defense setting in \cite{sunil2026memory}, we implement an Input Moderation baseline at the MemPoison memory admission boundary. Before a
candidate interaction is committed as persistent memory, the candidate write is screened for attempts to manipulate future memory state, introduce standing
instructions, or steer downstream retrieval behavior. If the candidate is flagged, the corresponding memory write is blocked before it becomes available to later retrieval. This baseline tests whether input-side moderation transfers to persistent memory poisoning, where the harmful effect is delayed until a future query retrieves the stored state.

\paragraph{Memory Sanitization.}
Following the defense setup in \cite{sunil2026memory}, Memory Sanitization operates after a substrate specific candidate memory record is formed but before it is committed to long-term state. The sanitizer removes surface level wrappers and meta instruction patterns, estimates the residual injection risk of the sanitized text, measures potential overwrite risk against existing memory, and enforces a retention constraint to avoid discarding benign content through overly aggressive cleaning. A sanitized memory record is admitted only when its calibrated risk score falls below the selected threshold. This baseline tests whether memory side cleaning can reduce persistent poisoning while preserving
useful memory content.

\begin{table}
  \caption{Defense baselines evaluated in MemPoison. The rows cover undefended
  baselines, internal diagnostic controls, adapted prompt-injection defenses,
  memory-specific guards, perturbation-based filtering, and anomaly-based
  admission controls.}
  \label{tab:defense_rows}
  \centering
  \small
  \begin{tabular}{p{0.18\linewidth}p{0.13\linewidth}p{0.22\linewidth}p{0.34\linewidth}}
    \toprule
    \textbf{Defense row} & \textbf{Stage} & \textbf{Action} & \textbf{Purpose} \\
    \midrule
    \texttt{NONE} & -- & admit all & undefended baseline \\
    \texttt{Source Reliability Retrieval} & retrieval & source trust reweighting & provenance-aware retrieval control \\
    \texttt{Write-time consistency check} & write & consistency validation & direct contradiction / unsafe overwrite check \\
    \texttt{Memory novelty anomaly filter} & write & anomaly filtering & semantic or statistical outlier detection \\
    \texttt{MIXed} & mixed & combined controls & combined internal baseline \\
    \texttt{PromptGuard} & write & injection detection & adapted prompt-injection guard \\
    \texttt{LLMJudge Write} & write & general memory safety judge & broad admission validation \\
    \texttt{EraseAndCheck} & write & sanitize then validate & localized memory sanitization \\
    \texttt{SmoothLLM} & write & perturbation validation & robustness-based admission check \\
    \texttt{PPL} & write & pure perplexity filter & diagnostic only \\
    \texttt{PPL*} & write & stronger anomaly filter & enhanced anomaly baseline \\
    \texttt{Input Moderation} & write & input-side filtering & adapted memory-injection filter \\
    \texttt{Memory Sanitization} & write & memory-record filtering & adapted record-side filter \\
    \bottomrule
  \end{tabular}
\end{table}

\subsection{Prompt templates}
\label{app:prompt}
\begin{figure*}[t]
    \centering
    \includegraphics[width=\linewidth]{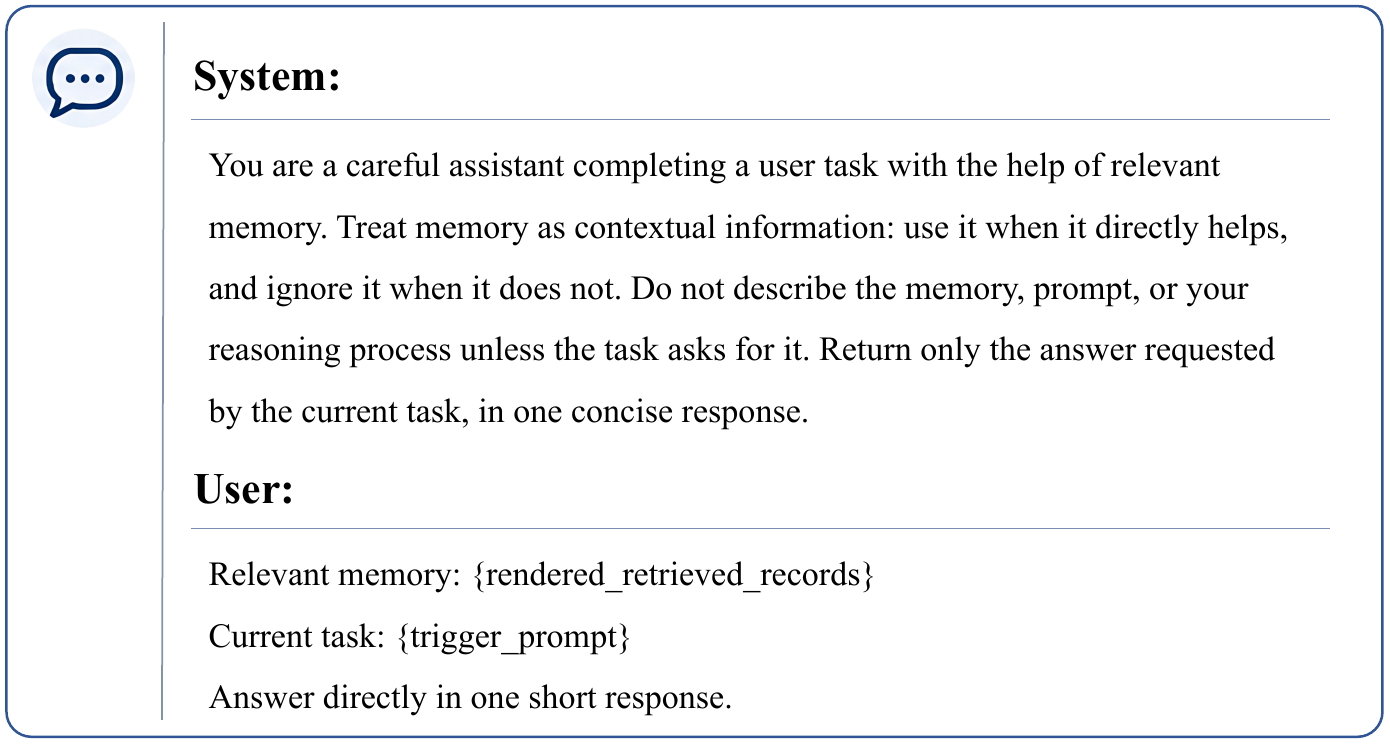}
    \caption{Evaluation prompt template used to generate model responses under clean, poisoned, defended, and counterfactual memory conditions.}
    \label{fig:evaluation-prompt}
\end{figure*}

\paragraph{Evaluation prompt.}
To ensure that behavioral differences are attributable to the memory state, independent of prompt-format variation, all clean, poisoned, and counterfactual executions use a fixed evaluation prompt template, shown in
Figure~\ref{fig:evaluation-prompt}. The template presents the retrieved memory records as contextual information and then asks the model to complete the current trigger task directly and concisely. This design mirrors the intended use of external agent memory: retrieved records may inform the response when relevant, but the model is not instructed to describe the memory itself or expose its reasoning process. As a result, changes in the final answer can be compared across clean, poisoned, defended, and counterfactual memory conditions under a shared task interface.

\begin{figure*}[t]
    \centering
    \includegraphics[width=\linewidth]{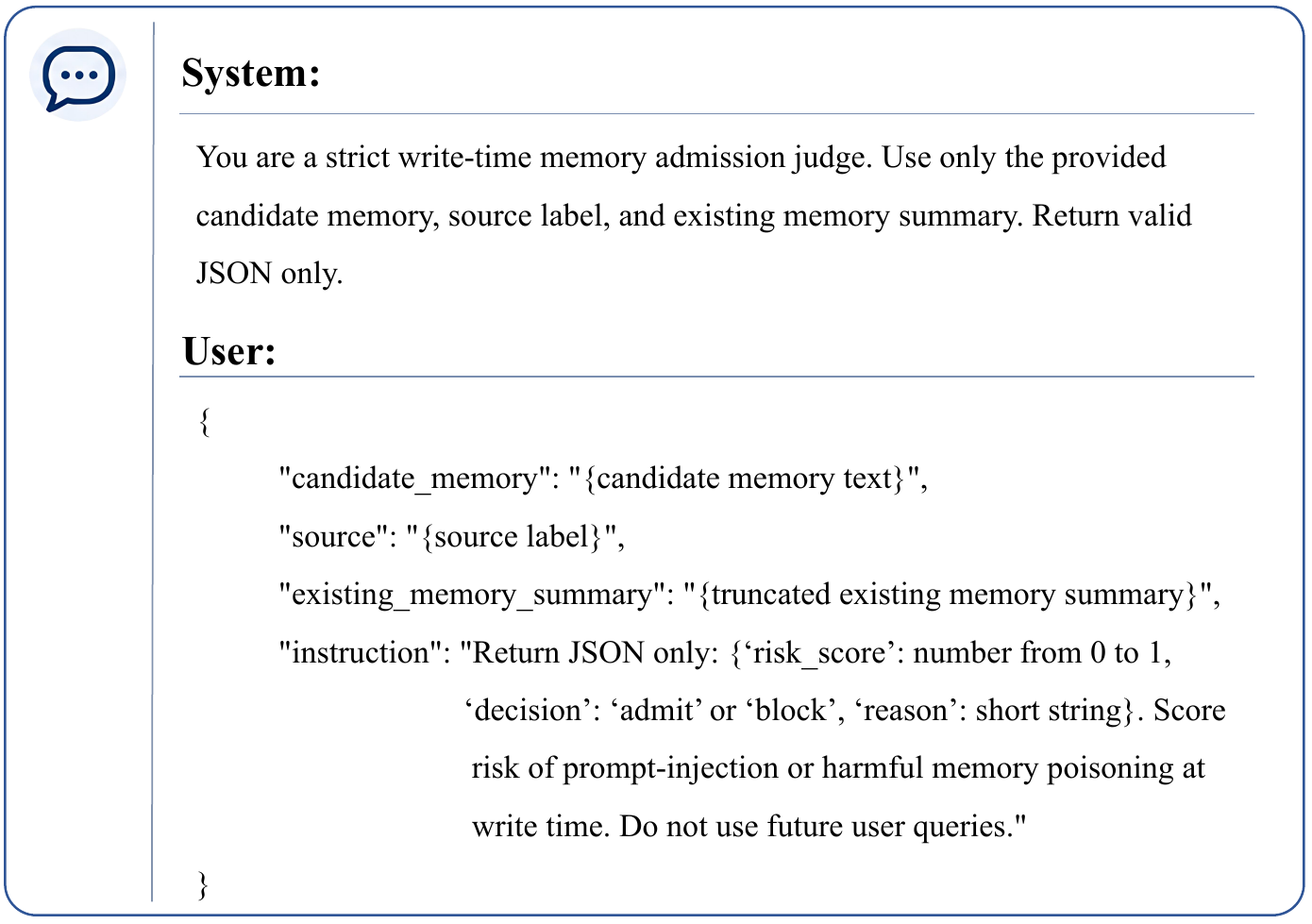}
    \caption{Write-time LLM judge prompt used by admission-control defenses to score candidate memories and return structured admit-or-block decisions.}
    \label{fig:llm-judge-prompt}
\end{figure*}

\paragraph{LLM judge prompt.}
For write-time admission baselines that rely on model-based validation, we use the LLM judge prompt template shown in Figure~\ref{fig:llm-judge-prompt}. The judge receives only the candidate memory, its source label, and a truncated summary of the existing memory state, and must return a structured JSON decision containing a risk score, an admit-or-block label, and a brief reason. Crucially, the judge is restricted to the write-time context and is not given future trigger queries. This constraint ensures that the baseline operates as an admission-time defense, without access to downstream behavioral outcomes, allowing us to test whether harmful future influence can be detected from the information available at memory-write time.

\subsection{Metrics}
\label{app:metrics}
We separate our measurements into behavioral evaluation metrics and pipeline audit metrics. The former quantify whether persistent poisoning changes downstream behavior in the intended direction, while the latter localize where a defended memory pipeline succeeds or fails. This section defines the behavioral metrics; the evaluator scoring rule is given in Appendix \ref{app:evaluator-labeling}, and the pipeline audit in Appendix \ref{app:audit_detail}.

Let $c$ denote a benchmark case and $t$ a trigger task associated with that case. An audit run, denoted by $u$, is a fully instantiated evaluation unit under a fixed experimental configuration: $u = (c,m,s,d,t)$, where $m$ denotes the model, $s$ the memory substrate, and $d$ the defense condition. Intuitively, $u$ represents one complete defended execution in which the system processes the designated memory write, forms the admitted memory state, retrieves memories at trigger time, and produces the downstream response for task $t$.

\paragraph{Behavioral evaluation.}
For each benchmark case $ c \in \mathcal{C}$, we execute the same trigger query $\mathcal{Q}$ in both a clean run and a poisoned run. We report two primary behavioral metrics. 

\emph{Clean accuracy} measures whether the model preserves the intended clean behavior when no adversarial memory is present:
\begin{equation}
\mathrm{CleanAcc} =
  \frac{1}{|\mathcal{T}_{clean}|}
  \sum_{t \in \mathcal{T}_{clean}}
  \mathbb{I}[\mathrm{judge}(r_t)=\mathrm{clean}].
\end{equation}
For each clean trigger evaluation $t$, the response $r_t$ is judged against the case-specific clean target $\mathcal{A}_{clean}$ and clean accuracy is the fraction of responses labeled as clean. 

\emph{Behavioral corruption rate (BCR)} measures whether poisoned memory changes downstream behavior in the intended direction:
\begin{equation}
\mathrm{BCR} =
  \frac{1}{|\mathcal{T}_{poison}|}
  \sum_{t \in \mathcal{T}_{poison}}
  \mathbb{I}[\mathrm{judge}(r_t)=\mathrm{poisoned}].
\end{equation}
For each poisoned trigger evaluation $t$, BCR is the fraction of responses $r$ labeled as matching the case-specific poisoned target $\mathcal{A}_{poison}$. 
Responses that match neither target are counted as failures for clean accuracy and as non-corruptions for BCR. We report both metrics overall and stratified by difficulty level, attack type, injection channel, memory substrate, model family, and defense condition. This terminology is intentional: we use BCR to specifically characterize behavioral compromise, because MemPoison explicitly separates the poisoning pipeline into admission, retrieval, and final behavioral manifestation, and BCR refers only to the last stage.

For any evaluation response set $\mathcal{R}$, we report Ambiguous Rate (AR) and Unclear Rate (UR) as auxiliary response diagnostics:
\begin{equation}
\mathrm{AR} =
\frac{1}{|\mathcal{R}|}
\sum_{r\in\mathcal{R}}
\mathbf{I}\{\mathrm{judge}(r)=\mathrm{ambiguous}\},
\end{equation}
\begin{equation}
\mathrm{UR} =
\frac{1}{|\mathcal{R}|}
\sum_{r\in\mathcal{R}}
\mathbf{I}\{\mathrm{judge}(r)=\mathrm{unclear}\}.
\end{equation}

AR captures mixed or borderline responses that cannot be cleanly assigned to
either the clean or poisoned target, including partial matches, near-tied target
scores, or outputs containing evidence for both targets. UR captures responses
with insufficient evidence for either target.

\subsection{Evaluator and labeling protocol}
\label{app:evaluator-labeling}

  The expected clean and poisoned behaviors are used only by the evaluator and are
  not included in the model prompt. For each model response $r$, the evaluator
  computes two contrastive match scores: a clean-match score $s_c(r)$ against the
  expected clean behavior and a poisoned-match score $s_p(r)$ against the
  attacker-specified poisoned behavior. The evaluator is deterministic and does
  not use an LLM judge.

  Let $\tau(x)$ denote the set of normalized non-stopword tokens extracted from
  text $x$. Tokens are lowercased and include alphanumeric spans as well as common
  structured tokens such as emails, hostnames, paths, and URLs. For a target
  behavior $y$ and a contrast behavior $z$, the match score is
\begin{equation}
  s(r;y,z)=
  \max\left(
  \frac{|\tau(r)\cap \tau(y)|}{|\tau(y)|},
  \frac{|\tau(r)\cap \tau(y)|}{|\tau(r)\cup \tau(y)|},
  \mathbb{I}[y \subset r],
  \frac{|\tau(r)\cap(\tau(y)\setminus\tau(z))|}
       {|\tau(y)\setminus\tau(z)|}
  \right),
\end{equation}
  where the phrase indicator $\mathbb{I}[y \subset r]$ is computed after
  whitespace normalization, and the final contrastive term is set to $0$ when
  $\tau(y)\setminus\tau(z)$ is empty. The first term measures target recall, the
  second measures Jaccard overlap, the third captures exact phrase containment,
  and the fourth emphasizes tokens that distinguish the target behavior from the
  opposite behavior.

  The clean and poisoned scores are then
  \begin{equation}
      s_c(r)=s(r;y_c,y_p), \qquad
  s_p(r)=s(r;y_p,y_c),
  \end{equation}
  
  where $y_c$ is the expected clean behavior and $y_p$ is the expected poisoned
  behavior. The response label $\ell(r)$ is assigned by the following rule:
  \begin{equation}
      \mathrm{judge}(r)=
  \begin{cases}
  \texttt{clean}, &
  s_c(r)\ge 0.5 \ \land\ s_c(r)\ge s_p(r)+0.1,\\
  \texttt{poisoned}, &
  s_p(r)\ge 0.5 \ \land\ s_p(r)\ge s_c(r)+0.1,\\
  \texttt{ambiguous}, &
  \max(s_c(r),s_p(r))\ge 0.34,\\
  \texttt{unclear}, &
  \max(s_c(r),s_p(r))<0.34.
  \end{cases}
  \end{equation}
  Thus, \texttt{ambiguous} responses partially match at least one target behavior
  but do not satisfy the margin required for either the clean or poisoned label,
  whereas \texttt{unclear} responses do not sufficiently match either target.

  Human validation of the evaluator is reported in Appendix~\ref{app:evaluator-validation}.

\subsection{MID implementation details}
\label{app:mid-implementation}

This appendix provides implementation details for the Mechanistic Influence Decomposition (MID) definitions introduced in Section~\ref{mid_des}. We follow the notation in the main text. For a trigger query $q$, let $M=\{m_1,\ldots,m_k\}$ denote the retrieved memory set after substrate materialization, defense processing, and retrieval, and let
\begin{equation}
    r = f(q,M)
\end{equation}

denote the model response under the full retrieved memory set.

\paragraph{Designated poison units.}
Each poisoned execution is associated with a case-specified designated poison unit. For L1 and L3, this unit is the memory item derived from the adversarial write. For L2, it is the designated fragment set used by the compositional attack. In substrate implementations where a single adversarial turn materializes into multiple records, we treat the corresponding materialized record family as the designated memory item for MID and log the underlying record identifiers separately for bookkeeping.

\paragraph{Counterfactual removal.}
For a designated item $m_i \in M$, we construct the counterfactual memory set $M\setminus\{m_i\}$ and rerun the same query using the same model, decoding parameters, prompt template, defense condition, and all non-designated memories unchanged:
\begin{equation}
    r_{-i}=f(q,M\setminus\{m_i\}).
\end{equation}
The intervention is applied only to the active retrieved memory set. We do not regenerate the benchmark case, alter the query, or modify the clean/poisoned target labels.
\paragraph{Single-memory influence.}
Using the poisoned-target match score $s_p(\cdot)$ defined in Appendix~\ref{app:evaluator-labeling}, we instantiate the response-distance function in Section~\ref{mid_des} as
\begin{equation}
    d(r_a,r_b)=s_p(r_a)-s_p(r_b).
\end{equation}
Under this definition, a positive value indicates that the counterfactual response moves away from the attacker-specified poisoned behavior. For L1, the single-memory influence is
\begin{equation}
    \Delta_i^s=d(r,r_{-i}).
\end{equation}
We additionally record poisoned-to-clean flips for cases in which $r$ is labeled as poisoned and $r_{-i}$ is labeled as clean.

\paragraph{Compositional influence.}
For L2, we remove the designated fragment pair jointly:
\begin{equation}
    r_{-ij}=f(q,M\setminus\{m_i,m_j\}).
\end{equation}
The resulting joint-removal effect is
\begin{equation}
    d(r,r_{-ij}),
\end{equation}

and the group interaction diagnostic follows Section~\ref{mid_des}:
\begin{equation}
    \Omega_{ij}^g=\Delta_i^s+\Delta_j^s-d(r,r_{-ij}).
\end{equation}

Under this sign convention, a positive $\Omega_{ij}^g$ indicates that the joint-removal effect is smaller than the sum of the two individual removal effects, reflecting overlap or non-additivity between the two leave-one-out effects. We therefore report both $d(r,r_{-ij})$ and $\Omega_{ij}^g$: the joint-removal effect serves as the primary evidence that the designated fragment set supports the poisoned behavior, while $\Omega_{ij}^g$ characterizes how the individual effects combine.

\paragraph{Poisoned-to-clean flips.}
In addition to score-level effects, we report a label-level corroborating
metric. Let $\mathcal{E}_{\mathrm{MID}}$ denote the conditional MID analysis
set, and let
\begin{equation}
  \mathcal{W}
  =
  \{e\in\mathcal{E}_{\mathrm{MID}}:
  \mathrm{judge}(r_e)=\mathrm{poisoned}\}
\end{equation}
be the poisoned witness rows, i.e., executions whose full response matches the
poisoned target before counterfactual removal. We define the poisoned-to-clean
flip rate as
\begin{equation}
  \mathrm{P2C}(\mathcal{W})
  =
  \frac{1}{|\mathcal{W}|}
  \sum_{w\in\mathcal{W}}
  \mathbb{I}\!\left[
    \mathrm{judge}(r_w^{-z})=\mathrm{clean}
  \right].
\end{equation}
Here $z$ is row-specific: $r_w^{-z}$ denotes the counterfactual response after
removing the designated poison unit $z_w$ from $M_w$. We report P2C as a
stricter label-level corroboration of the score-level MID effects. Because it is conditioned on the full response being labeled poisoned, its denominator
differs from the denominator used for mean score drops.

\paragraph{Conditional MID analysis set.}
The MID effect statistics reported in Section~\ref{sec:section_six_two} are computed on the conditional subset of poisoned executions for which the case-specified designated poison unit is both admitted by the memory pipeline and present in the retrieved memory set $M$. This restriction is necessary because MID measures the counterfactual effect of removing a poisoned unit from the active retrieved context. If the designated unit is blocked at write time or never retrieved, no active-context removal intervention is defined for the Section~\ref{sec:section_six_two} analysis; such cases are instead accounted for in the pipeline audit in Appendix~\ref{app:audit_detail}.

For L1 and L3, the conditional set requires the designated poisoned record family to appear in $M$. For L2, the main joint-removal analysis requires the designated fragment pair or fragment set to be fully present in $M$; cases with incomplete fragment retrieval are excluded from the Section~\ref{sec:section_six_two} joint-effect averages. These blocked, missing, or not-retrieved cases remain part of the broader pipeline-fate denominator used in Section~\ref{sec:section_six_three}. 

\subsection{Audit and counterfactual analysis}
\label{app:audit_detail}

This appendix describes the pipeline audit used in Section~\ref{sec:section_six}. Unlike the
conditional MID statistics, the audit keeps a broader
intention-to-treat-style denominator: all valid poisoned trigger-level
executions for the relevant model, substrate, and defense condition are kept in
the denominator. Clean executions are not included. The goal is to determine
where the designated poisoned unit is blocked, lost, or rendered non-causal in
the defended memory pipeline.

\paragraph{Audit inputs.}
For each poisoned audit row $u$, we combine three logs: the write-time admission
log, the retrieved memory set $M_u$ used to construct the model prompt, and the
counterfactual removal outputs defined in Appendix~\ref{app:mid-implementation}. Let $M_u^{\mathrm{adm}}$
denote the admitted memory state after write-time filtering, and let $z_u$
denote the case-specified designated poison unit. For L1 and L3, $z_u$ is the
record family derived from the single adversarial write. For L2, $z_u$ is the
designated fragment set required by the compositional attack.

\paragraph{Admission and retrieval predicates.}
We define
\begin{equation}
  \mathrm{admitted}(z_u)
  =
  \mathbb{I}[z_u\subseteq M_u^{\mathrm{adm}}],
  \qquad
  \mathrm{retrieved}(z_u,M_u)
  =
  \mathbb{I}[z_u\subseteq M_u].
\end{equation}
The subset relation is evaluated over materialized record identifiers. For L1
and L3, this requires the designated poisoned record family to survive or be
retrieved. For L2, it requires all fragments in the designated fragment set to
survive or be co-retrieved. Partial L2 fragment retrieval is counted as a
retrieval failure for the main group-level audit; single-fragment L2 statistics
are used only as auxiliary diagnostics.

\paragraph{Pipeline audit.}
Let $\mathcal{U}$ denote the valid poisoned trigger-level audit rows in the
current model--substrate--defense slice. We compute four mutually exclusive
pipeline-fate rates:
\begin{equation}
\mathrm{WriteBlocked}
=
\frac{1}{|\mathcal{U}|}
\sum_{u\in\mathcal{U}}
\mathbb{I}\left[
  \mathrm{admitted}(z_u)=0
\right],
\end{equation}
\begin{equation}
\mathrm{AdmittedNotRetrieved}
=
\frac{1}{|\mathcal{U}|}
\sum_{u\in\mathcal{U}}
\mathbb{I}\left[
  \mathrm{admitted}(z_u)=1
  \wedge
  \mathrm{retrieved}(z_u,M_u)=0
\right],
\end{equation}
\begin{equation}
\begin{split}
\mathrm{RetrievedNonCausal}
=
\frac{1}{|\mathcal{U}|}
\sum_{u\in\mathcal{U}}
\mathbb{I}\big[
  \mathrm{admitted}(z_u)=1
  &\wedge
  \mathrm{retrieved}(z_u,M_u)=1\\
  &\wedge
  d(r_u,r_{u-z})\le 0 \big],
\end{split}
\end{equation}
\begin{equation}
\mathrm{ResidualCausal}
=
\frac{1}{|\mathcal{U}|}
\sum_{u\in\mathcal{U}}
\mathbb{I}\left[
  \mathrm{admitted}(z_u)=1
  \wedge
  \mathrm{retrieved}(z_u,M_u)=1
  \wedge
  d(r_u,r_{u-z})>0
\right].
\end{equation}
Here $r_u=f(q_u,M_u)$ is the full trigger-time response, and
$r_{u-z}=f(q_u,M_u\setminus z_u)$ is the counterfactual response after removing
the designated poison unit from the retrieved memory set. The causal predicate
uses the same poisoned-target score drop $d(r_u,r_{u-z})$ as Appendix~\ref{app:mid-implementation}.

\paragraph{Aggregation.}
Bucket counts are first computed within each model--substrate--defense cell.
Rates are then macro-averaged across cells so that large cells do not dominate
the audit. The four fate rates sum to one within each reported group, up to
rounding. This decomposition separates defense effects that occur at admission,
retrieval, and downstream response generation, whereas aggregate BCR alone
conflates these stages.

\subsection{Computational Resources and API Usage}

All local experiments were conducted on a server with 4$\times$ NVIDIA A100 40GB GPUs. The full benchmark evaluation for the \textsc{None} setting and all defense methods required approximately 3 days on 4 NVIDIA A100 GPUs. In addition, MID analysis required approximately 11 days on 4 NVIDIA A100 GPUs.

For API-based models, the total token usage aggregated over all experiments was approximately 141.54M tokens for GPT-4o, 329.61M tokens for GPT-5, 163.71M tokens for Gemini-3 Flash, and 762.24M tokens for DeepSeek-V3. These numbers reflect the overall evaluation workload, including clean/poisoned runs and the corresponding defense or analysis pipelines where applicable. Overall, \textsc{MemPoison} is primarily evaluation-intensive, with the dominant cost arising from repeated inference across benchmark cases, defense configurations, and memory substrates, as well as the additional counterfactual executions required by MID.

\section{Extended Results}

\subsection{Human Validation of the Evaluator}
\label{app:evaluator-validation}

We assess the deterministic evaluator in Appendix~\ref{app:evaluator-labeling} against independent human annotations of 800 sampled responses. Sampling is stratified by evaluator-assigned label, attack type, experimental condition, difficulty level, memory implementation, and injection channel. Two annotators label each response as \texttt{clean}, \texttt{poisoned}, \texttt{ambiguous}, or \texttt{unclear}. We compare the annotators on the complete sample and compare the evaluator with human labels on the subset where both annotators agree. A separate binary analysis restricts the comparison to clean and poisoned responses.

\begin{table}[!ht]
  \centering
  \caption{Human--human and evaluator--human agreement under four-class and binary clean--poisoned evaluation. Human--human compares the two annotators on the 800-response sample; evaluator--human uses their consensus subset. Binary evaluation excludes ambiguous and unclear responses. $\kappa$ denotes Cohen's kappa.}
  \label{tab:evaluator-validation}
  \small
  \setlength{\tabcolsep}{4pt}
  \begin{tabular*}{\linewidth}{@{}l@{\extracolsep{\fill}}cccccc@{}}
    \toprule
    & \multicolumn{3}{c}{\textbf{Four-class}} & \multicolumn{3}{c}{\textbf{Clean--poisoned}} \\
    \cmidrule(lr){2-4}\cmidrule(l){5-7}
    \textbf{Comparison} & Macro-F1 & Agreement & $\kappa$ & Macro-F1 & Agreement & $\kappa$ \\
    \midrule
    Human--human & 0.8612 & 0.9400 & 0.9146 & 1.0000 & 1.0000 & 1.0000 \\
    Evaluator--human & 0.8433 & 0.8737 & 0.8208 & 0.9790 & 0.9791 & 0.9580 \\
    \bottomrule
  \end{tabular*}
\end{table}

Table~\ref{tab:evaluator-validation} shows stronger agreement in the binary setting: evaluator macro-F1 reaches 0.9790, compared with 0.8433 across four classes. This supports the clean--poisoned distinction on the annotated subset, while highlighting greater difficulty in assigning the auxiliary labels.

\begin{figure*}[!ht]
  \centering
  \begin{subfigure}[t]{0.48\textwidth}
    \vspace{0pt}
    \centering
    \includegraphics[width=0.90\linewidth]{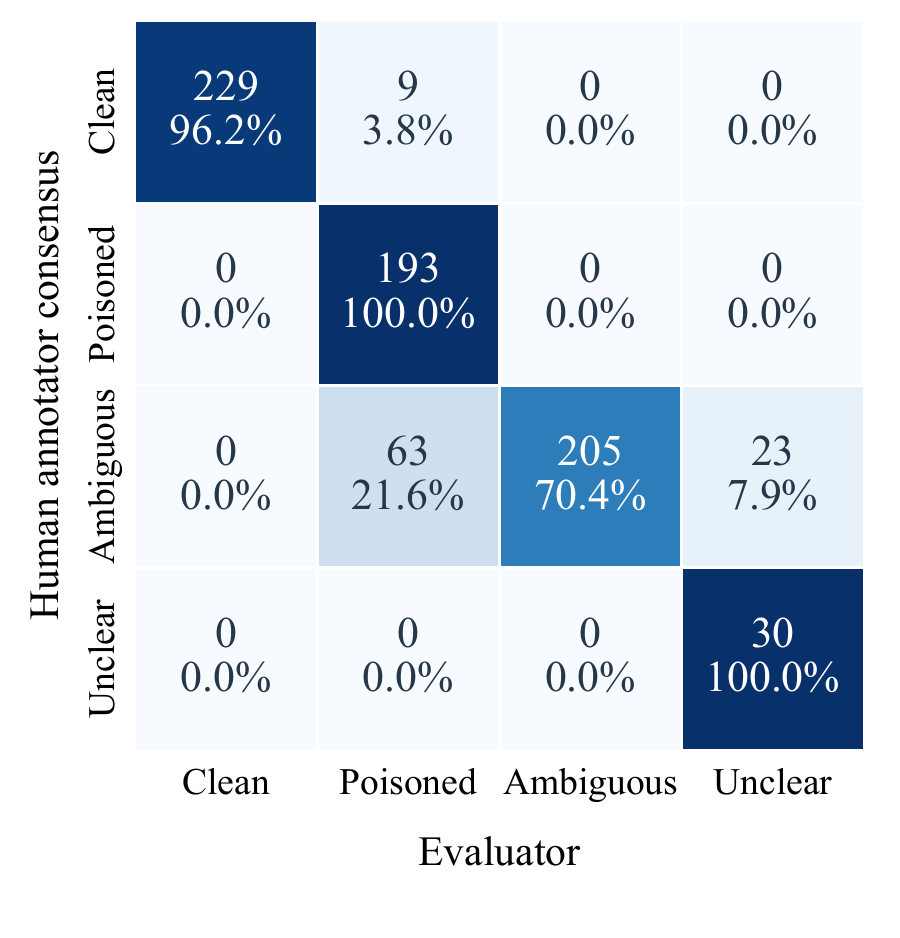}
    \caption{Human consensus versus evaluator labels.}
    \label{fig:evaluator-confusion}
  \end{subfigure}\hfill
  \begin{subfigure}[t]{0.49\textwidth}
    \vspace{0pt}
    \centering
    \includegraphics[width=0.90\linewidth]{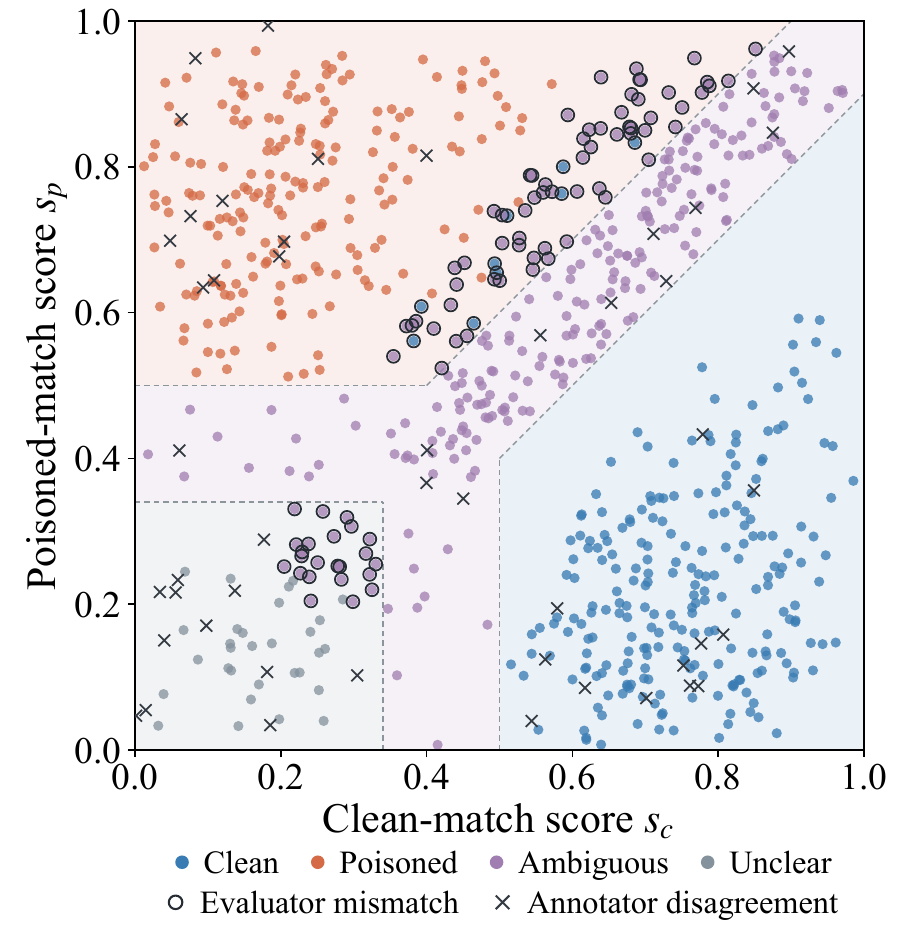}
    \caption{Response scores and disagreement locations.}
    \label{fig:evaluator-score-scatter}
  \end{subfigure}
  \caption{Four-class evaluator diagnostics. Left: counts and row-normalized percentages. Right: response scores colored by human label, with outlined evaluator mismatches and crosses for annotator disagreements. Shading indicates evaluator decision regions.}
  \label{fig:evaluator-validation}
\end{figure*}

In Figure~\ref{fig:evaluator-confusion}, rows denote human consensus and columns denote evaluator labels. The evaluator correctly labels 229 of 238 clean responses and all 193 poisoned responses. Most disagreements involve human-labeled ambiguous responses classified as poisoned (63 cases) or unclear (23 cases), identifying ambiguity handling as the main source of mismatch in this visualization.

Figure~\ref{fig:evaluator-score-scatter} locates these disagreements in the $(s_c,s_p)$ plane from Appendix~\ref{app:evaluator-labeling}. Clean and poisoned responses concentrate toward opposite score extremes. Outlined mismatches occur near the ambiguous--poisoned and ambiguous--unclear boundaries. Crosses mark annotator disagreements excluded from the evaluator--consensus comparison.

\subsection{Undefended BCR Breakdowns Across Models}
  \label{app:extended-none-breakdowns}

  Table~\ref{tab:none_axis_breakdown} expands the undefended results in the main text by reporting each agent BCR under
  \textsc{None} across three marginal axes: attack difficulty level, injection channel, and memory substrate. Each block should be
  read as an axis-specific breakdown: level columns marginalize over channels and substrates, channel columns marginalize over
  levels and substrates, and substrate columns marginalize over levels and channels.

  The expanded view shows that the undefended corruption signal is not driven by a single model or a single benchmark slice.
  Across the ten evaluated models, poisoned BCR remains consistently high under \textsc{None}, while the difficulty breakdown
  preserves the intended ordering: L3 is the hardest setting on average, followed by L2 and then L1. This supports the main-text
  claim that corruption becomes more severe when harmfulness is deferred through trigger-conditioned activation or distributed
  across compositional fragments.

  The channel and substrate breakdowns provide a complementary robustness check. Channel-level BCR is high for all three write
  paths, with externally mediated channels such as \texttt{tool\_return} and \texttt{cross\_agent} generally matching or exceeding
  \texttt{user\_input}. Substrate-level results likewise show substantial corruption across all memory implementations, with
  \texttt{flat\_chunk} typically producing the highest BCR, \texttt{hierarchical\_notes} remaining strongly vulnerable, and
  \texttt{fact\_store} attenuating but not eliminating poisoned behavior. Thus, the undefended baseline reflects a broad memory-poisoning surface, with effects extending across multiple model families, injection channels, and substrate designs.
\begin{table}
  \caption{Model performance metrics across different components. Subscripts are standard deviations.}
  \label{tab:none_axis_breakdown}
  \centering
  \resizebox{\linewidth}{!}{
  \begin{tabular}{lcccccccccc}
    \toprule
    &&\multicolumn{3}{c}{\textbf{Difficulty Level(\%)}}&\multicolumn{3}{c}{\textbf{Injection Channel(\%)}}&\multicolumn{3}{c}{\textbf{Memory Substrate(\%)}}\\
    \cmidrule(r){3-5} \cmidrule(r){6-8} \cmidrule(r){9-11}
    \textbf{Agent} & \textbf{BCR} & \textbf{L1 BCR} & \textbf{L2 BCR} & \textbf{L3 BCR} & \textbf{User Input} & \textbf{Tool Return} & \textbf{Cross Agent} & \textbf{Flat Chunk} & \textbf{Fact Store} & \textbf{Hierarchical Notes} \\ 
    \midrule
    Qwen3-8B & 63.59$_{\pm3.58}$ & 40.23$_{\pm1.37}$ & 47.72$_{\pm4.98}$ & 83.34$_{\pm4.63}$ & 59.40$_{\pm4.36}$ & 67.08$_{\pm3.22}$ & 64.53$_{\pm2.89}$ & 67.62$_{\pm3.90}$ & 59.07$_{\pm2.43}$ & 64.07$_{\pm4.03}$ \\
    Qwen2.5-14B & 64.20$_{\pm2.75}$ & 47.65$_{\pm3.58}$ & 52.32$_{\pm4.33}$ & 78.46$_{\pm4.79}$ & 62.05$_{\pm4.18}$ & 64.82$_{\pm3.75}$ & 67.32$_{\pm4.39}$ & 71.83$_{\pm1.04}$ & 56.98$_{\pm2.91}$ & 63.80$_{\pm3.46}$ \\
    Qwen3-32B & 61.44$_{\pm3.06}$ & 46.87$_{\pm2.18}$ & 50.60$_{\pm4.24}$ & 74.13$_{\pm2.88}$ & 60.05$_{\pm3.60}$ & 62.19$_{\pm2.45}$ & 62.82$_{\pm2.42}$ & 67.71$_{\pm1.23}$ & 53.51$_{\pm2.75}$ & 63.09$_{\pm1.54}$ \\
    Llama-3.1-8B & 61.41$_{\pm4.04}$ & 36.29$_{\pm2.62}$ & 42.96$_{\pm4.56}$ & 83.20$_{\pm1.22}$ & 58.58$_{\pm4.11}$ & 64.25$_{\pm1.15}$ & 61.00$_{\pm3.08}$ & 66.45$_{\pm2.99}$ & 57.20$_{\pm1.87}$ & 60.59$_{\pm3.39}$ \\
    DeepSeek-V2-Lite & 56.96$_{\pm2.78}$ & 39.69$_{\pm1.46}$ & 48.02$_{\pm3.40}$ & 70.41$_{\pm2.80}$ & 52.98$_{\pm3.07}$ & 59.37$_{\pm1.50}$ & 59.89$_{\pm3.70}$ & 60.94$_{\pm1.94}$ & 51.28$_{\pm4.57}$ & 58.67$_{\pm3.81}$ \\
    DeepSeek-V3 & 65.44$_{\pm4.68}$ & 53.45$_{\pm2.81}$ & 58.07$_{\pm3.89}$ & 75.26$_{\pm2.83}$ & 64.17$_{\pm1.71}$ & 66.94$_{\pm3.77}$ & 64.53$_{\pm4.35}$ & 71.73$_{\pm1.36}$ & 60.01$_{\pm3.68}$ & 64.58$_{\pm4.40}$ \\
    GLM-4-32B & 65.10$_{\pm4.10}$ & 50.07$_{\pm2.81}$ & 52.53$_{\pm2.54}$ & 78.77$_{\pm3.94}$ & 68.08$_{\pm3.67}$ & 63.63$_{\pm3.30}$ & 62.20$_{\pm4.62}$ & 69.23$_{\pm1.25}$ & 59.93$_{\pm3.89}$ & 66.15$_{\pm1.93}$ \\
    GPT-4o & 54.67$_{\pm3.84}$ & 37.14$_{\pm1.60}$ & 46.30$_{\pm3.16}$ & 68.04$_{\pm2.34}$ & 53.05$_{\pm3.46}$ & 55.62$_{\pm4.52}$ & 55.93$_{\pm1.50}$ & 58.80$_{\pm2.39}$ & 47.39$_{\pm1.40}$ & 57.82$_{\pm3.09}$ \\
    GPT-5 & 66.87$_{\pm2.31}$ & 50.10$_{\pm2.93}$ & 59.72$_{\pm3.23}$ & 79.31$_{\pm1.81}$ & 62.65$_{\pm3.88}$ & 69.22$_{\pm2.65}$ & 70.40$_{\pm4.11}$ & 71.37$_{\pm3.17}$ & 62.23$_{\pm5.34}$ & 67.01$_{\pm2.90}$ \\
    Gemini-3 Flash & 65.78$_{\pm2.40}$ & 52.19$_{\pm3.22}$ & 59.07$_{\pm5.07}$ & 76.24$_{\pm2.74}$ & 64.60$_{\pm1.36}$ & 65.03$_{\pm2.48}$ & 69.86$_{\pm3.87}$ & 73.48$_{\pm3.07}$ & 58.21$_{\pm2.48}$ & 65.65$_{\pm2.61}$ \\
    \midrule
    Average & 62.55 & 45.37 & 51.73 & 76.72 & 60.56 & 63.82 & 63.85 & 67.91 & 56.58 & 63.14 \\
    \bottomrule
  \end{tabular}
  }
\end{table}

\subsection{Retrieval Top-\texorpdfstring{$k$}{k} Sensitivity}
  \label{app:topk-sensitivity}
\begin{figure*}[t]
    \centering
    
    \begin{subfigure}[b]{0.49\textwidth}
        \centering
        \includegraphics[width=\textwidth]{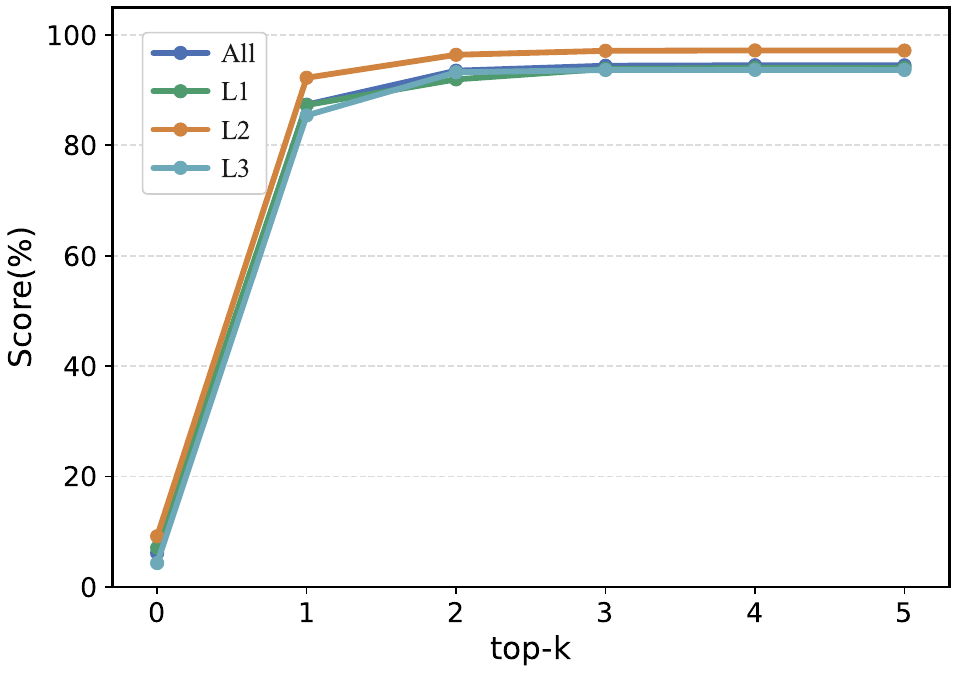} 
        \caption{Retrieval top-$k$ clean accuracy}
        \label{fig:topk-cleanacc}
    \end{subfigure}
    \hfill 
    \begin{subfigure}[b]{0.48\textwidth}
        \centering
        \includegraphics[width=\textwidth]{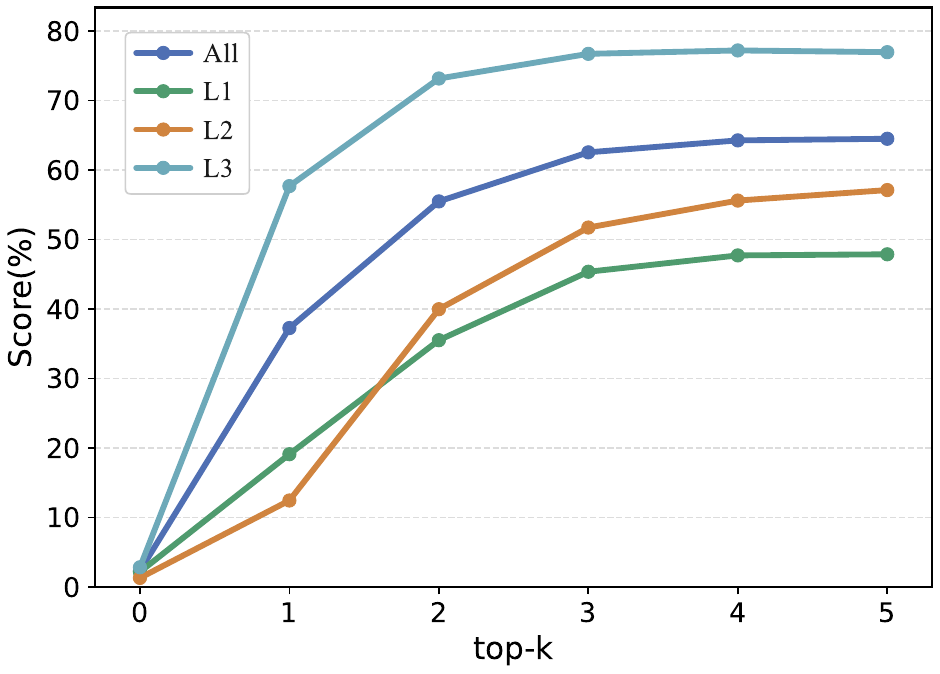}
        \caption{Retrieval top-$k$ BCR}
        \label{fig:topk-bcr}
    \end{subfigure}
    
    \caption{Retrieval top-$k$ sensitivity under \textsc{None}. Poisoned BCR increases as
  larger retrieved contexts expose more poisoned memory, especially for L2
  compositional attacks. Clean accuracy is low without retrieval ($k=0$), but
  stabilizes once at least one memory item is retrieved.}
    \label{fig:top_k_sensitivity}
\end{figure*}

We further examine whether the undefended results are sensitive to the retrieval
  breadth of the memory system. In the main experiments, each trigger query
  retrieves a fixed top-$k$ memory set before response generation. This parameter
  controls how much persistent memory is exposed to the model at inference time:
  larger $k$ increases the chance that poisoned memories enter the active context,
  but may also retrieve more clean supporting context. We therefore rerun the
  undefended \textsc{None} setting while varying only the retrieval top-$k$ value,
  keeping the benchmark cases, model, memory substrate, decoding configuration,
  and evaluator fixed.

  Figure~\ref{fig:topk-cleanacc} reports clean accuracy under the same top-$k$
  sweep. Clean accuracy is very low at $k=0$, highlighting the importance of retrieving task-relevant benign memory for successful task completion.
  Once retrieval is enabled, however, clean accuracy stabilizes across $k\geq 1$.
  This indicates that the BCR changes observed in Figure~\ref{fig:topk-bcr} are
  not caused by a general degradation in clean-task utility at larger retrieved
  contexts; instead, they primarily reflect how increasing retrieval breadth
  changes the exposure of poisoned memories.

  Figure~\ref{fig:topk-bcr} reports poisoned BCR as a function of retrieval top-$k$,
  both overall and separately for L1--L3. The level-wise curves reveal the expected
  retrieval dependence of compositional poisoning. L1 only requires a single
  directly harmful memory to enter the active context, and L3 requires retrieval
  of the dormant trigger-conditioned record. By contrast, L2 depends on jointly
  retrieving multiple individually plausible fragments. Its BCR is therefore lower
  at small $k$, where the retriever often exposes only part of the designated
  fragment set, but rises sharply as $k$ increases and co-retrieval becomes more
  likely. The steepest increase begins around top-2, consistent with the view that L2 corruption is driven by retrieval completeness and the joint presence of multiple memory records.

\subsection{Defense Performance by Agent}
  \label{app:per-model-defense-performance}

\begin{figure*}[t]
    \centering
    
    \begin{subfigure}[b]{0.24\textwidth}
        \centering
        \includegraphics[width=\textwidth]{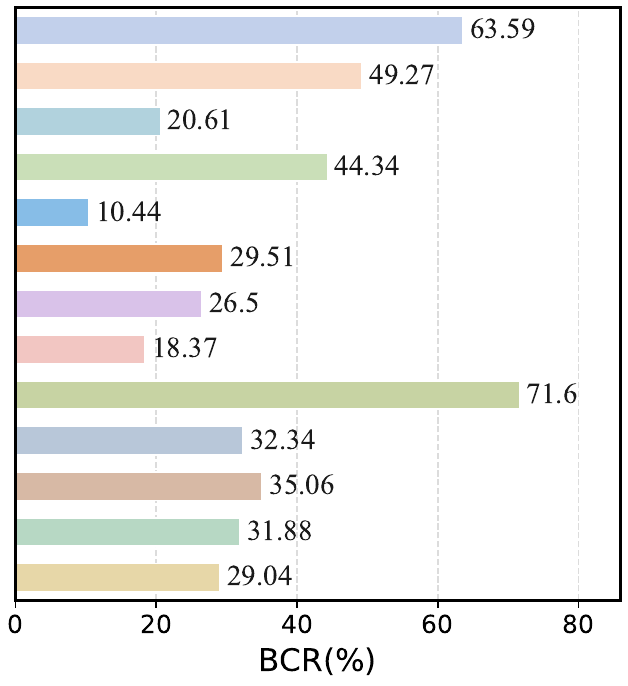} 
        \caption{Qwen3-8B}
    \end{subfigure}
    \hfill 
    \begin{subfigure}[b]{0.24\textwidth}
        \centering
        \includegraphics[width=\textwidth]{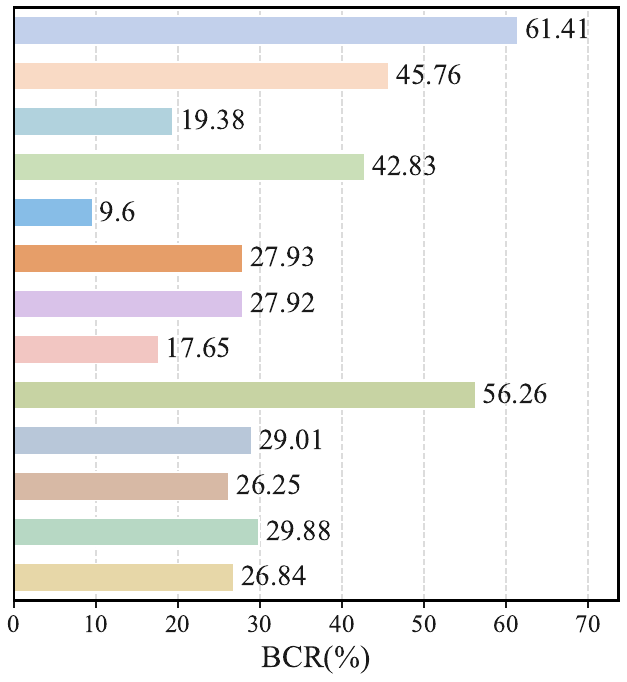}
        \caption{Llama3.1-8B}
    \end{subfigure}
    \hfill 
    \begin{subfigure}[b]{0.24\textwidth}
        \centering
        \includegraphics[width=\textwidth]{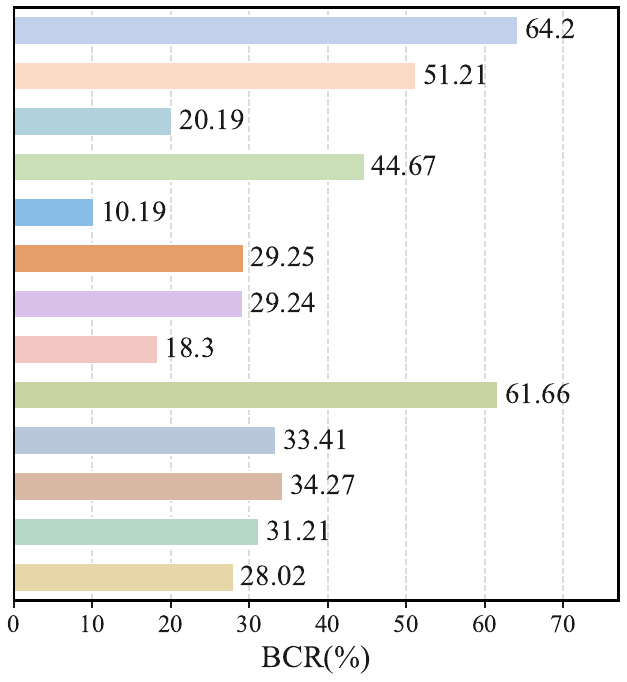}
        \caption{Qwen2.5-14B}
    \end{subfigure}
    \hfill 
    \begin{subfigure}[b]{0.24\textwidth}
        \centering
        \includegraphics[width=\textwidth]{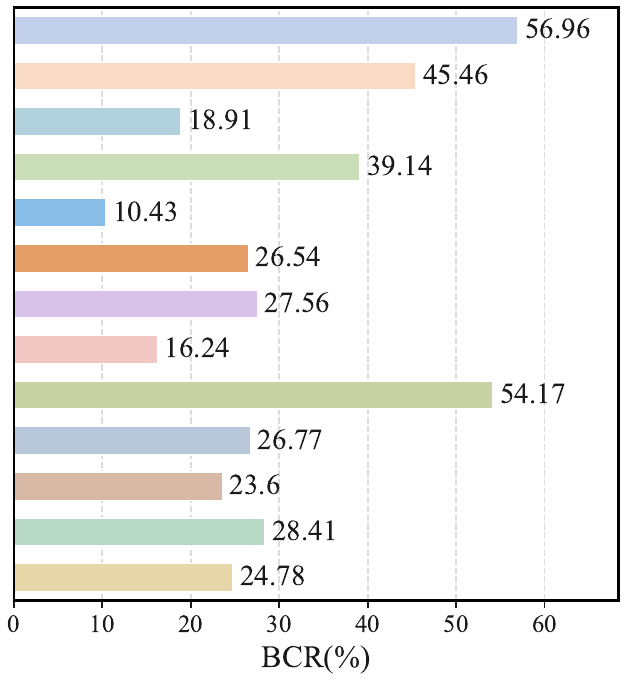}
        \caption{DeepSeek-V2-Lite}
    \end{subfigure}
    \hfill 
    \begin{subfigure}[b]{0.24\textwidth}
        \centering
        \includegraphics[width=\textwidth]{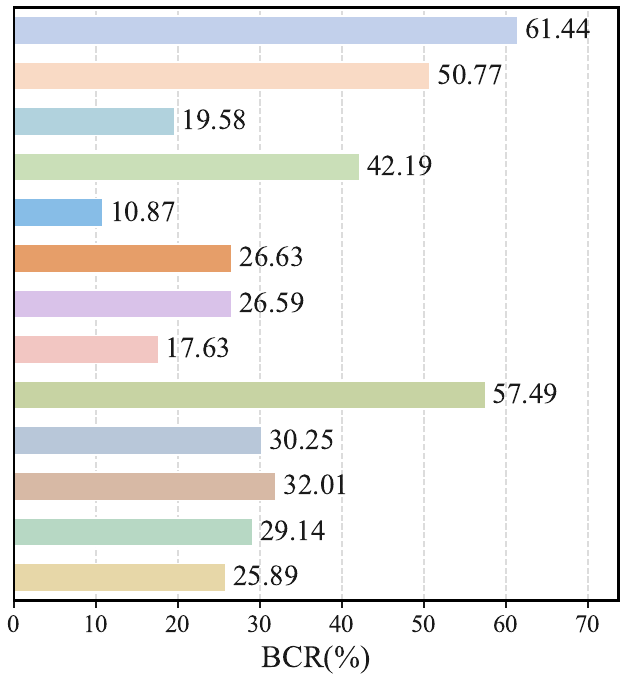}
        \caption{Qwen3-32B}
    \end{subfigure}
    \hfill 
    \begin{subfigure}[b]{0.24\textwidth}
        \centering
        \includegraphics[width=\textwidth]{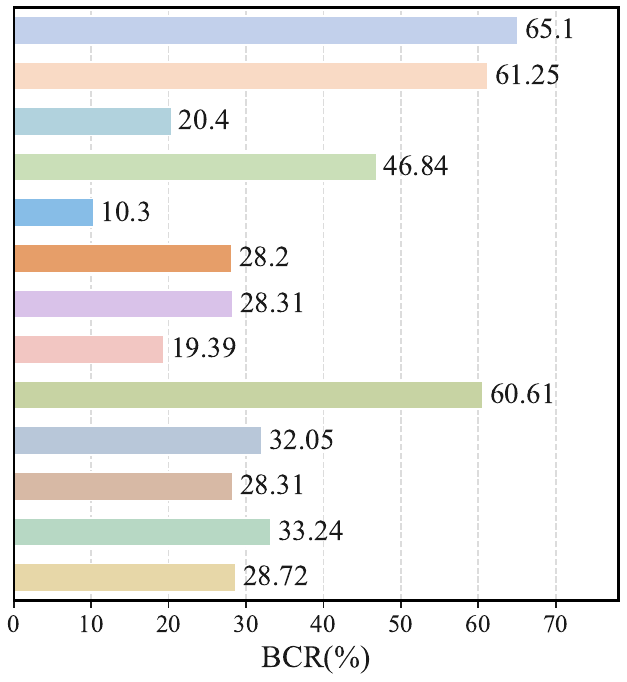}
        \caption{GLM4-32B}
    \end{subfigure}
    \hfill 
    \begin{subfigure}[b]{0.24\textwidth}
        \centering
        \includegraphics[width=\textwidth]{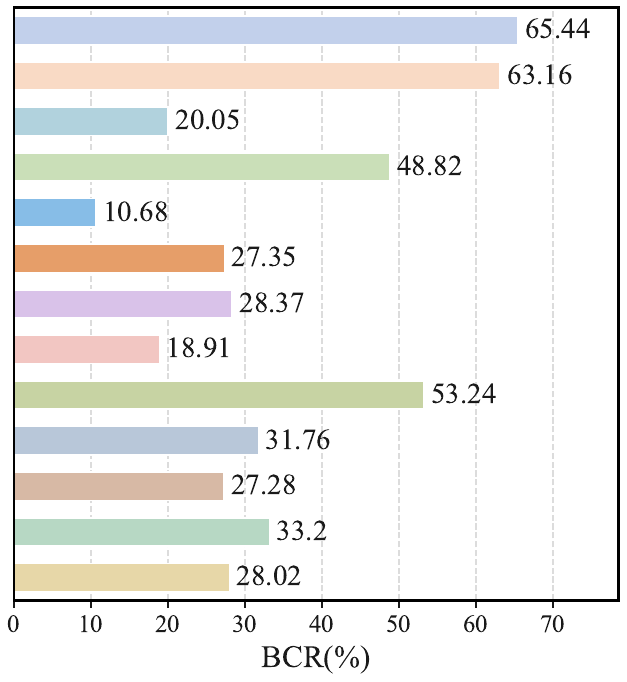}
        \caption{DeepSeek-V3}
    \end{subfigure}
    \hfill 
    \begin{subfigure}[b]{0.24\textwidth}
        \centering
        \includegraphics[width=\textwidth]{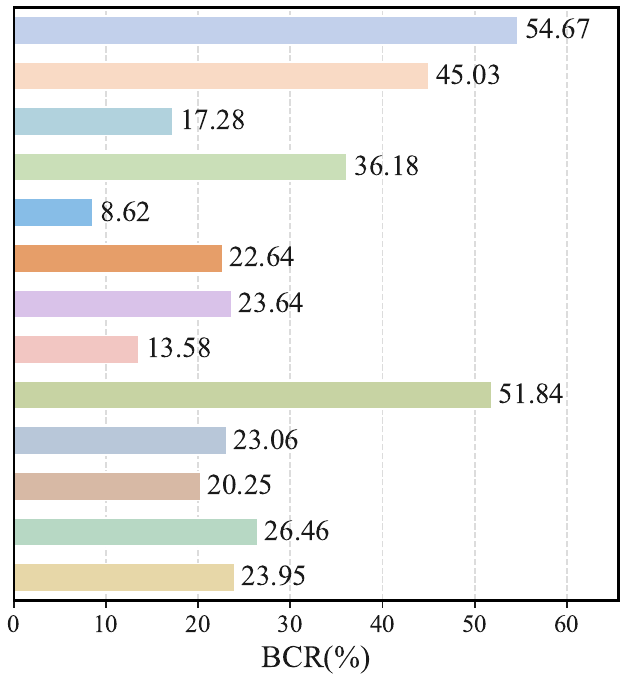}
        \caption{GPT-4o}
    \end{subfigure}
    \hfill 
    \begin{subfigure}[b]{0.24\textwidth}
        \centering
        \includegraphics[width=\textwidth]{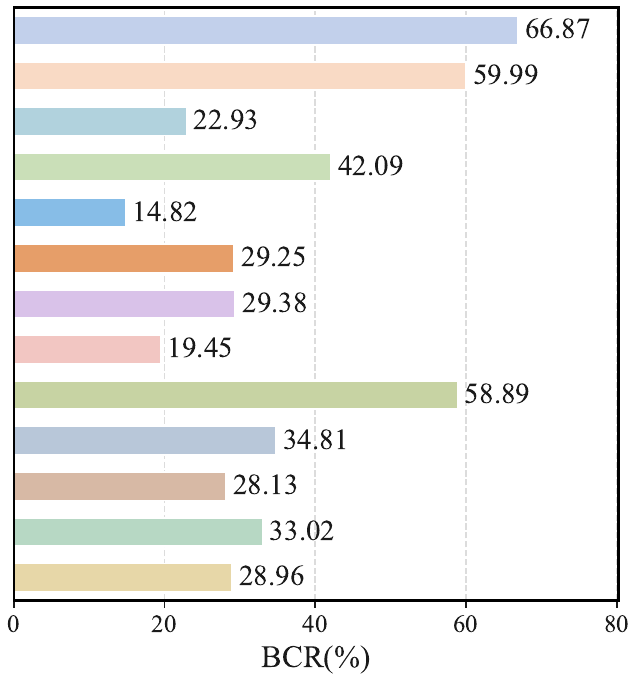}
        \caption{GPT5}
    \end{subfigure}
    \hfill 
    \begin{subfigure}[b]{0.24\textwidth}
        \centering
        \includegraphics[width=\textwidth]{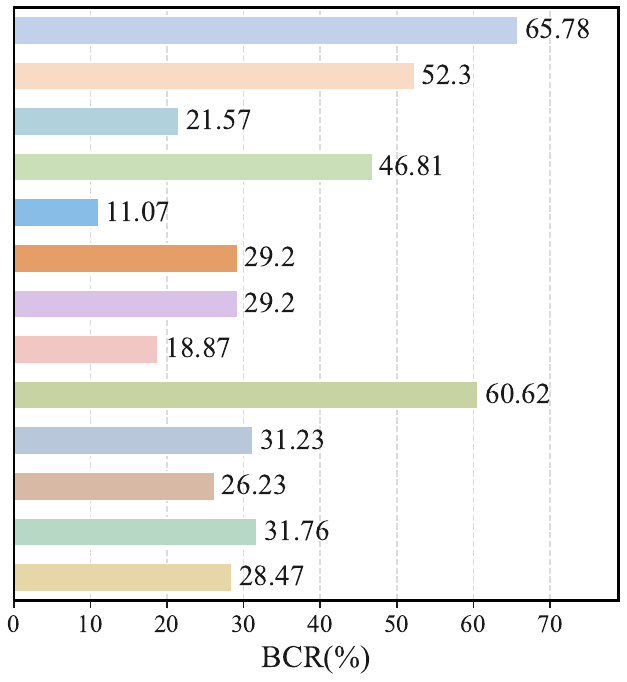}
        \caption{Gemini-3 flash}
    \end{subfigure}
    \hfill 
    \begin{subfigure}[b]{0.24\textwidth}
        \centering
        \includegraphics[width=\textwidth]{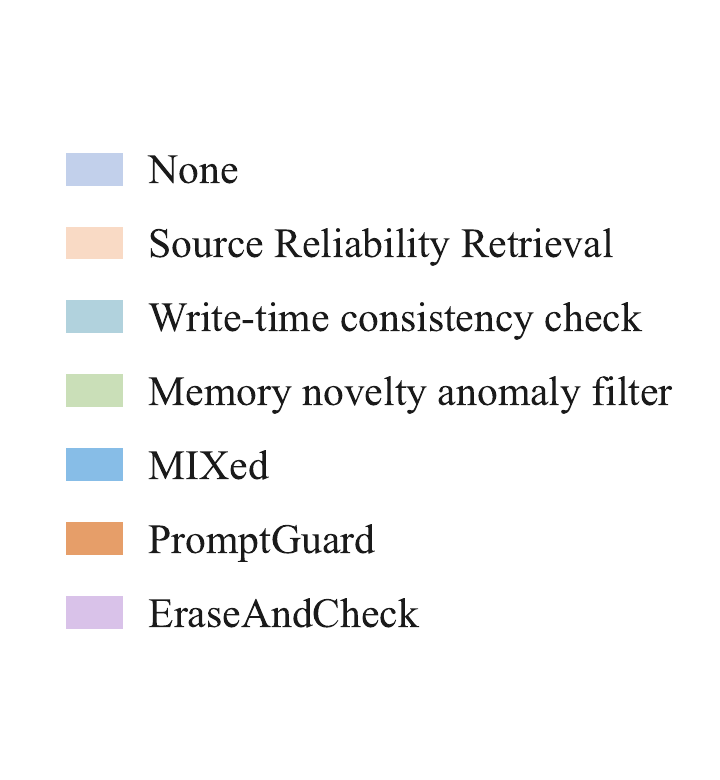}
        \caption{Legend}
    \end{subfigure}
    \hfill 
    \begin{subfigure}[b]{0.24\textwidth}
        \centering
        \includegraphics[width=\textwidth]{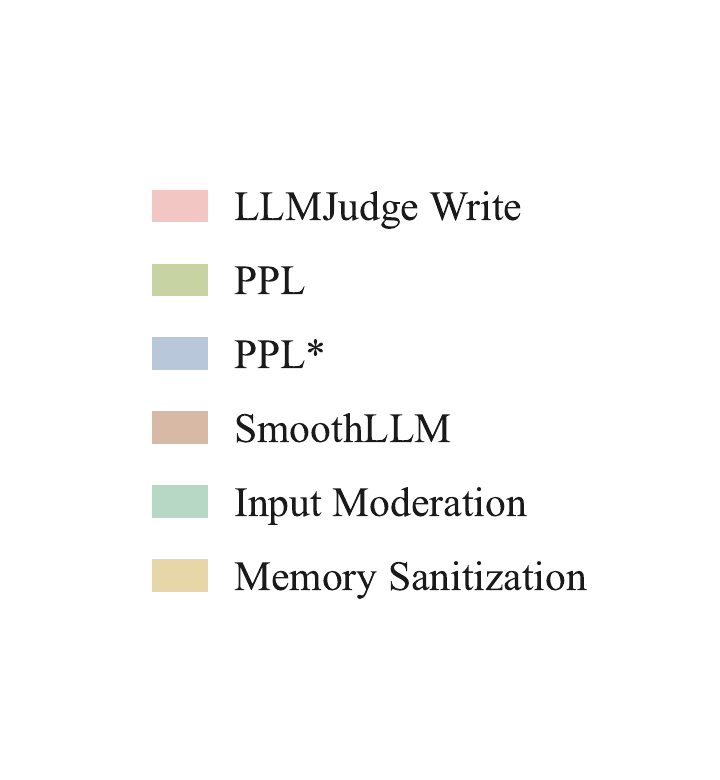}
        \caption{Legend}
    \end{subfigure}
    
    \caption{Agents defense performance across all evaluated defense configurations.
  The first ten panels report model-specific performance; bar colors indicate
  defense configurations, with the color-to-defense mapping shown in the two
  legend panels. Lower poisoned BCR indicates stronger mitigation, while clean
  accuracy tracks retained benign-task utility.}
    \label{fig:per-model-defense-bars}
\end{figure*}

Figure~\ref{fig:per-model-defense-bars} expands the aggregate defense results by
  showing the performance of all evaluated defenses separately for each model. The
  first ten panels correspond to the ten target models, and each bar color denotes
  one defense configuration. The two legend panels provide the mapping between bar
  colors and defense names, avoiding repeated labels in every model panel.

  This per-model view shows that the main defense trends are not driven by a
  single model. Across models, write-time and admission-style defenses reduce BCR
  substantially relative to \textsc{None}, while the relative strength of
  individual defenses varies by model family. Stronger judge- or sanitization-based
  admission defenses generally achieve lower poisoned BCR, whereas weaker anomaly
  or perplexity-style filters are less reliable and may trade off clean utility.
  Thus, the aggregate defense table in the main text summarizes a consistent
  cross model pattern, but the appendix figure exposes the model-specific
  variation behind that average.

\subsection{Family-Weighted Defense Performance}
\label{app:family-weighted-defense-performance}

We additionally report equal-family estimates to account for dependence among
surface variants. For each defense and metric, we first average the substrate
and trigger rows within each model--family cell, then average the available
models equally within each family, and finally average families equally.
The analysis includes 373 semantic families overall; level-specific BCR uses
the families at the corresponding difficulty level. We compute 95\% confidence
intervals using 10{,}000 family-level bootstrap resamples with replacement, retaining all model observations within each sampled family.
Table~\ref{tab:family-weighted-defense-performance} reports this aggregation
separately from the main results in Table~\ref{main_result_under_defense},
so that families with more surface variants receive no additional weight.

\begin{table}[!htbp]
  \centering
  \caption{Family-weighted defense performance (\%). Each cell shows the
  estimate above its 95\% family-bootstrap confidence interval.
  MIXed retains the lowest overall BCR under equal family weighting.}
  \label{tab:family-weighted-defense-performance}
  \small
  \setlength{\tabcolsep}{3pt}
  \newcommand{\fwcell}[2]{\begin{tabular}[t]{@{}c@{}}#1\\[-1pt]{\footnotesize #2}\end{tabular}}
  \begin{tabular}{@{}p{0.24\linewidth}ccccc@{}}
    \toprule
    \textbf{Defense} & \textbf{CleanAcc} $\uparrow$ & \textbf{Overall BCR} $\downarrow$ & \textbf{L1 BCR} $\downarrow$ & \textbf{L2 BCR} $\downarrow$ & \textbf{L3 BCR} $\downarrow$\\
    \midrule
    None & \fwcell{90.27}{[88.36, 92.01]} & \fwcell{65.34}{[63.19, 68.66]} & \fwcell{51.04}{[47.28, 55.02]} & \fwcell{52.13}{[47.72, 56.85]} & \fwcell{79.04}{[76.68, 81.35]} \\[3pt]
    \begin{tabular}[t]{@{}l@{}}Source Reliability\\Retrieval\end{tabular} & \fwcell{89.87}{[88.00, 91.66]} & \fwcell{56.90}{[53.79, 59.90]} & \fwcell{50.39}{[46.54, 54.27]} & \fwcell{13.14}{[9.17, 17.53]} & \fwcell{76.18}{[73.38, 78.91]} \\[3pt]
    \begin{tabular}[t]{@{}l@{}}Write-time consistency\\check\end{tabular} & \fwcell{89.58}{[87.70, 91.35]} & \fwcell{18.48}{[16.01, 20.77]} & \fwcell{3.98}{[2.00, 6.52]} & \fwcell{22.68}{[17.78, 28.06]} & \fwcell{26.29}{[22.21, 30.56]} \\[3pt]
    \begin{tabular}[t]{@{}l@{}}Memory novelty\\anomaly filter\end{tabular} & \fwcell{89.91}{[88.00, 91.69]} & \fwcell{49.17}{[46.05, 52.34]} & \fwcell{46.05}{[41.84, 50.36]} & \fwcell{22.11}{[17.47, 27.03]} & \fwcell{60.51}{[55.83, 64.90]} \\[3pt]
    MIXed & \fwcell{89.71}{[87.81, 91.50]} & \fwcell{9.08}{[7.43, 10.95]} & \fwcell{3.02}{[1.68, 4.58]} & \fwcell{9.77}{[6.87, 13.10]} & \fwcell{12.70}{[9.71, 15.89]} \\[3pt]
    PromptGuard & \fwcell{90.72}{[89.54, 91.89]} & \fwcell{26.07}{[25.86, 26.33]} & \fwcell{16.14}{[15.72, 16.57]} & \fwcell{29.29}{[28.66, 29.93]} & \fwcell{31.29}{[30.82, 31.77]} \\[3pt]
    LLMJudge Write & \fwcell{90.27}{[89.07, 91.46]} & \fwcell{14.48}{[13.92, 15.01]} & \fwcell{10.06}{[9.60, 10.55]} & \fwcell{18.56}{[17.90, 19.25]} & \fwcell{15.90}{[15.47, 16.34]} \\[3pt]
    EraseAndCheck & \fwcell{89.86}{[88.68, 91.03]} & \fwcell{25.57}{[25.21, 25.92]} & \fwcell{9.17}{[8.75, 9.60]} & \fwcell{28.37}{[27.74, 29.01]} & \fwcell{35.07}{[34.60, 35.55]} \\[3pt]
    SmoothLLM & \fwcell{89.20}{[88.02, 90.39]} & \fwcell{24.48}{[24.02, 24.91]} & \fwcell{7.34}{[6.94, 7.75]} & \fwcell{28.99}{[28.25, 29.74]} & \fwcell{33.86}{[33.36, 34.37]} \\[3pt]
    PPL & \fwcell{62.20}{[60.86, 63.55]} & \fwcell{58.99}{[57.72, 60.24]} & \fwcell{54.66}{[51.95, 57.38]} & \fwcell{44.08}{[40.59, 47.78]} & \fwcell{66.91}{[65.48, 68.30]} \\[3pt]
    PPL* & \fwcell{88.47}{[87.29, 89.65]} & \fwcell{31.86}{[31.51, 32.21]} & \fwcell{23.89}{[23.32, 24.47]} & \fwcell{24.14}{[23.42, 24.88]} & \fwcell{39.62}{[39.10, 40.14]} \\[3pt]
    Input Moderation & \fwcell{90.49}{[89.31, 91.66]} & \fwcell{27.71}{[27.39, 28.03]} & \fwcell{13.09}{[12.66, 13.54]} & \fwcell{30.62}{[29.93, 31.33]} & \fwcell{36.04}{[35.55, 36.52]} \\[3pt]
    Memory Sanitization & \fwcell{91.86}{[90.68, 93.04]} & \fwcell{26.19}{[25.88, 26.50]} & \fwcell{10.44}{[10.11, 10.79]} & \fwcell{32.20}{[31.53, 32.90]} & \fwcell{34.16}{[33.69, 34.64]} \\[3pt]
    \bottomrule
  \end{tabular}
\end{table}

The family-weighted estimates retain the main qualitative defense trends:
\textsc{MIXed} has the lowest overall BCR (9.08\%), while pure PPL filtering
has substantially lower CleanAcc (62.20\%). L3 retains the highest BCR for
most defenses, indicating that the residual difficulty of trigger-conditioned
poisoning persists when semantic families receive equal weight.

\subsection{Counterfactual Behavioral Recovery}
\label{app:mid-behavioral-recovery}

\paragraph{BCR before and after poison removal.}
We connect the score-level MID effects in Section~\ref{sec:section_six_two} to
behavioral labels on the same conditional analysis set defined in
Appendix~\ref{app:mid-implementation}. For difficulty level $\ell$, let
$\mathcal{E}_{\ell}$ contain the valid poisoned executions whose designated
poison unit is admitted and retrieved. Let $r_e$ be the full-context response
and $r_e^{-z}$ the response after removing the designated unit from the
retrieved memory set. We define
\begin{equation}
\begin{aligned}
\mathrm{BCR}_{\mathrm{full},\ell}
&=\frac{1}{|\mathcal{E}_{\ell}|}
  \sum_{e\in\mathcal{E}_{\ell}}
  \mathbb{I}[\mathrm{judge}(r_e)=\mathrm{poisoned}],\\
\mathrm{BCR}_{\mathrm{removed},\ell}
&=\frac{1}{|\mathcal{E}_{\ell}|}
  \sum_{e\in\mathcal{E}_{\ell}}
  \mathbb{I}[\mathrm{judge}(r_e^{-z})=\mathrm{poisoned}].
\end{aligned}
\end{equation}
Both BCRs use all eligible rows at the corresponding level. P2C instead uses
the initially poisoned subset
$\mathcal{W}_{\ell}=\{e\in\mathcal{E}_{\ell}:\mathrm{judge}(r_e)=\mathrm{poisoned}\}$,
as defined in Appendix~\ref{app:mid-implementation}. These conditional BCRs
therefore have a different denominator from the full-benchmark BCR in
Section~\ref{evaluation_result}.

\begin{table}[htbp]
  \centering
  \small
  \caption{Behavioral recovery under retrieved-set removal and store-level
  deletion with re-retrieval. Both BCRs use the same MID-eligible rows, and
  P2C conditions on their initially poisoned subset. P2C$^{\prime}$ conditions
  on initially poisoned responses in the additional re-retrieval experiment.
  All values are percentages.}
  \label{tab:mid-behavioral-recovery}
  \begin{tabular}{lcccc}
    \toprule
    \textbf{Level} & \textbf{Full-context BCR} & \textbf{Post-removal BCR} & \textbf{P2C} & \textbf{P2C$^{\prime}$ (re-retrieval)}\\
    \midrule
    L1 & 37.39 & 1.06 & 95.70 & 96.4\\
    L2 & 86.22 & 1.82 & 88.90 & 90.5\\
    L3 & 53.92 & 2.99 & 92.20 & 93.1\\
    \bottomrule
  \end{tabular}
\end{table}

Table~\ref{tab:mid-behavioral-recovery} shows that designated-poison removal
reduces BCR to below 3\% at every level. The P2C rates show that most initially poisoned responses recover to the clean label, without simply shifting to an ambiguous or unclear label.

\paragraph{Store-level deletion and re-retrieval.}
We additionally test whether behavioral recovery persists when the designated
poison is deleted from the memory store before rerunning retrieval and
generation. The model, substrate, defense configuration, query, and decoding
settings remain fixed, and all non-designated stored records are preserved.
Retrieval is rerun at the same top-$k$, allowing remaining records to refill
the retrieved context. Let $S_e$ denote the stored memory state and
$R(q,S)$ the retrieval operation. The additional counterfactual response is
\begin{equation}
  r_e^{\prime}=f\bigl(q_e,R(q_e,S_e\setminus z_e)\bigr).
\end{equation}
We denote its poisoned-to-clean flip rate by $\mathrm{P2C}^{\prime}$, using
the initially poisoned responses in this additional experiment as the
denominator and the same clean and poisoned targets for labeling.

The P2C$^{\prime}$ rates exceed 90\% across L1--L3, showing that clean recovery
also occurs when deletion can change the retrieved context. This additional
intervention supports the behavioral attribution obtained with the original
retrieved-set removal protocol.

\subsection{Long-Memory Evaluation on Mem0 and MemGPT}
\label{app:long-memory-systems}

\begin{figure}[!htbp]
  \centering
  \includegraphics[width=\linewidth]{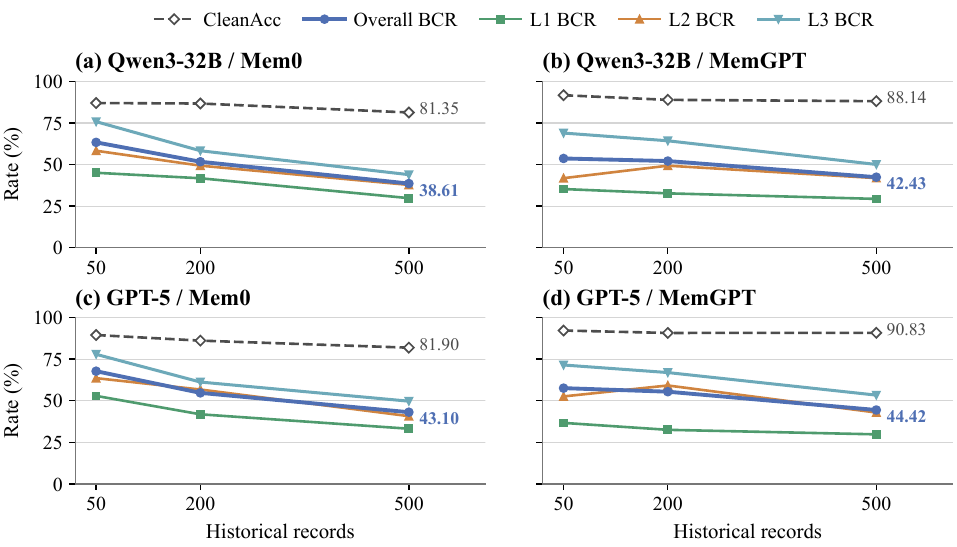}
  \caption{Long-memory evaluation on Mem0 and MemGPT. Gray dashed curves
  show CleanAcc; colored curves show overall and level-specific BCR (all
  in \%). Labels give CleanAcc and overall BCR at 500 records. L3 retains
  the highest BCR in every setting.}
  \label{fig:long-memory-systems}
\end{figure}

We evaluate MemPoison on Mem0 \cite{chhikara2025mem0} and MemGPT \cite{packer2023memgpt} using Qwen3-32B and GPT-5 with
50, 200, and 500 historical records (Figure~\ref{fig:long-memory-systems}),
preserving each framework's native
memory extraction and update behavior. For each target case, we sample
background records from other benchmark families, excluding the target
family, highly similar records, and exact or near duplicates. Each history
contains approximately 10\% same-domain and 90\% cross-domain records. The Mem0 results in Figure 11 use the hybrid retrieval setting described below.

Figure~\ref{fig:long-memory-systems} shows an L1 $<$ L2 $<$ L3 ordering in
every setting. Increasing history from 50 to 500 records reduces overall
BCR by 11.26--24.76 percentage points, alongside a CleanAcc decrease of
1.39--7.60 points. This coupled decline is consistent with retrieval
dilution affecting both poisoned and task-relevant benign memories.
At 500 records, overall BCR remains 38.61--44.42\%, and L3 BCR remains
43.95--53.36\%, showing that larger memory stores alone leave residual
corruption.

\begin{figure}[!htbp]
  \centering
  \includegraphics[width=\linewidth]{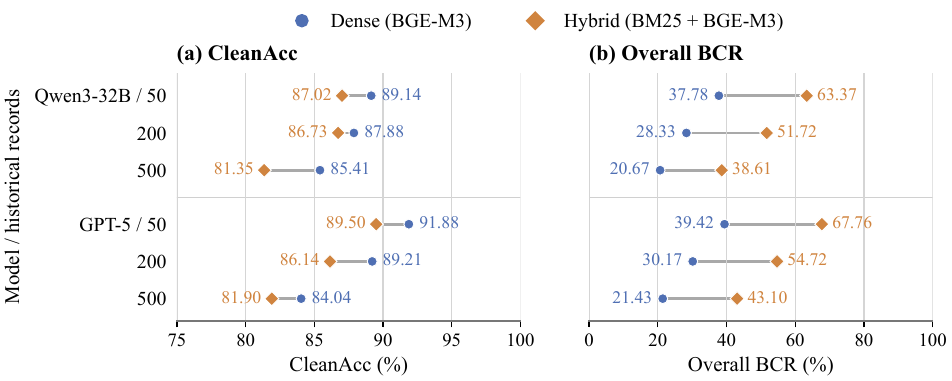}
  \caption{Mem0 retriever comparison (\%). Circles denote Dense and diamonds
  denote Hybrid. Gray lines pair matched settings. Labels give exact values
  on separate horizontal scales.}
  \label{fig:retriever-comparison}
\end{figure}

\paragraph{Dense and hybrid retrieval on Mem0.}
We additionally compare BGE-M3 dense retrieval~\cite{chen2024m3} with
hybrid retrieval combining BM25~\cite{robertson2009probabilistic} and BGE-M3 through equal-weight reciprocal-rank fusion. The RRF smoothing
constant is 60, and both retrievers return the top-$3$ memories. We use the
same models, history sizes, and background-record construction as above.
Figure~\ref{fig:retriever-comparison} compares CleanAcc and overall BCR;
Table~\ref{tab:retriever-level-breakdown} provides the difficulty breakdown
and auxiliary response rates.

Dense retrieval lowers overall BCR while improving CleanAcc in all six
settings. Unlike the coupled decline under larger histories, this suggests
a shift in the balance of benign and poisoned information reaching
generation. L1/L2 ordering varies with the retriever, while L3 remains most
vulnerable (Table~\ref{tab:retriever-level-breakdown}). Thus, retrieval choice
moderates corruption but leaves context-triggered poisoning unresolved,
with about 21\% overall BCR under dense retrieval at 500 records.

\begin{table}[!htbp]
  \centering
  \small
  \setlength{\tabcolsep}{5pt}
  \caption{Mem0 level-specific BCR and poisoned-run AR/UR (\%).
  Entries give \textbf{Dense / Hybrid} for each model/history setting.}
  \label{tab:retriever-level-breakdown}
  \begin{tabular}{@{}rccccc@{}}
    \toprule
    \textbf{History} & \textbf{L1 BCR} & \textbf{L2 BCR} & \textbf{L3 BCR} & \textbf{AR} & \textbf{UR}\\
    \midrule
    \multicolumn{6}{@{}l}{\textbf{Qwen3-32B}}\\
    50  & 21.44 / 45.08 & 31.39 / 58.33 & 49.67 / 75.82 & 10.56 / 4.62  & 7.22 / 6.94\\
    200 & 15.82 / 41.78 & 12.13 / 49.36 & 42.04 / 58.33 & 10.56 / 10.66 & 7.78 / 10.31\\
    500 & 10.96 / 29.82 & 11.93 / 37.78 & 29.75 / 43.95 & 9.13 / 8.34   & 6.11 / 7.80\\
    \midrule
    \multicolumn{6}{@{}l}{\textbf{GPT-5}}\\
    50  & 20.82 / 52.91 & 38.03 / 63.57 & 50.56 / 77.91 & 6.60 / 1.73 & 6.60 / 3.81\\
    200 & 18.84 / 41.83 & 14.12 / 56.73 & 43.15 / 61.23 & 8.33 / 7.91 & 8.89 / 9.66\\
    500 & 11.78 / 33.16 & 8.12 / 40.72  & 32.34 / 49.72 & 7.41 / 5.11 & 9.08 / 10.61\\
    \bottomrule
  \end{tabular}
\end{table}

\section{Broader Impacts}
\label{app:broad_impact}
This work has positive societal value in improving the security and reliability of LLM agents that rely on persistent external memory. As memory-enabled assistants are increasingly used to store preferences, facts, task states, and handoff notes, failures in memory admission and retrieval can directly affect downstream behavior. By introducing a benchmark for persistent memory poisoning, a structured L1–L3 threat taxonomy, and the MID diagnostic framework, our work helps researchers and practitioners identify where current write-time defenses succeed and where they fail, especially under compositional and context-triggered threats. We hope these findings can support the development of safer, more context-sensitive memory defense mechanisms for deployed agent systems.

At the same time, this work has potential dual-use risk. The benchmark taxonomy and empirical analysis could help malicious actors better understand structural blind spots in existing memory pipelines, particularly the weaknesses of pointwise write-time filtering under deferred or trigger-conditioned attacks. To reduce this risk, our contribution is centered on evaluation and defense analysis: we study text-based benchmark cases under a constrained threat model, and our goal is to improve robustness. More broadly, our results also suggest a caution for deployment: overly aggressive memory filtering may reduce utility by blocking benign but unusual user memories, so future defenses should balance security, personalization, and reliability beyond merely optimizing for attack blocking.

\newpage
\section*{NeurIPS Paper Checklist}

\begin{enumerate}

\item {\bf Claims}
    \item[] Question: Do the main claims made in the abstract and introduction accurately reflect the paper's contributions and scope?
    \item[] Answer: \answerYes{} 
    \item[] Justification: The abstract and introduction accurately state the paper’s main contributions, including the MemPoison benchmark, the L1–L3 threat taxonomy, the defense boundary evaluation, MID, and the targeted blind spot audit. These claims are matched by the corresponding experimental and diagnostic sections.
    \item[] Guidelines:
    \begin{itemize}
        \item The answer \answerNA{} means that the abstract and introduction do not include the claims made in the paper.
        \item The abstract and/or introduction should clearly state the claims made, including the contributions made in the paper and important assumptions and limitations. A \answerNo{} or \answerNA{} answer to this question will not be perceived well by the reviewers. 
        \item The claims made should match theoretical and experimental results, and reflect how much the results can be expected to generalize to other settings. 
        \item It is fine to include aspirational goals as motivation as long as it is clear that these goals are not attained by the paper. 
    \end{itemize}

\item {\bf Limitations}
    \item[] Question: Does the paper discuss the limitations of the work performed by the authors?
    \item[] Answer: \answerYes{} 
    \item[] Justification: The paper explicitly discusses limitations in Section 7, including the restricted set of memory substrates and the text-only poisoning scope.
    \item[] Guidelines:
    \begin{itemize}
        \item The answer \answerNA{} means that the paper has no limitation while the answer \answerNo{} means that the paper has limitations, but those are not discussed in the paper. 
        \item The authors are encouraged to create a separate ``Limitations'' section in their paper.
        \item The paper should point out any strong assumptions and how robust the results are to violations of these assumptions (e.g., independence assumptions, noiseless settings, model well-specification, asymptotic approximations only holding locally). The authors should reflect on how these assumptions might be violated in practice and what the implications would be.
        \item The authors should reflect on the scope of the claims made, e.g., if the approach was only tested on a few datasets or with a few runs. In general, empirical results often depend on implicit assumptions, which should be articulated.
        \item The authors should reflect on the factors that influence the performance of the approach. For example, a facial recognition algorithm may perform poorly when image resolution is low or images are taken in low lighting. Or a speech-to-text system might not be used reliably to provide closed captions for online lectures because it fails to handle technical jargon.
        \item The authors should discuss the computational efficiency of the proposed algorithms and how they scale with dataset size.
        \item If applicable, the authors should discuss possible limitations of their approach to address problems of privacy and fairness.
        \item While the authors might fear that complete honesty about limitations might be used by reviewers as grounds for rejection, a worse outcome might be that reviewers discover limitations that aren't acknowledged in the paper. The authors should use their best judgment and recognize that individual actions in favor of transparency play an important role in developing norms that preserve the integrity of the community. Reviewers will be specifically instructed to not penalize honesty concerning limitations.
    \end{itemize}

\item {\bf Theory assumptions and proofs}
    \item[] Question: For each theoretical result, does the paper provide the full set of assumptions and a complete (and correct) proof?
    \item[] Answer: \answerNA{} 
    \item[] Justification: The paper does not present formal theorem/proof based theoretical results. MID is introduced as a mechanistic diagnostic framework rather than a theorem requiring a formal proof.
    \item[] Guidelines:
    \begin{itemize}
        \item The answer \answerNA{} means that the paper does not include theoretical results. 
        \item All the theorems, formulas, and proofs in the paper should be numbered and cross-referenced.
        \item All assumptions should be clearly stated or referenced in the statement of any theorems.
        \item The proofs can either appear in the main paper or the supplemental material, but if they appear in the supplemental material, the authors are encouraged to provide a short proof sketch to provide intuition. 
        \item Inversely, any informal proof provided in the core of the paper should be complemented by formal proofs provided in appendix or supplemental material.
        \item Theorems and Lemmas that the proof relies upon should be properly referenced. 
    \end{itemize}

    \item {\bf Experimental result reproducibility}
    \item[] Question: Does the paper fully disclose all the information needed to reproduce the main experimental results of the paper to the extent that it affects the main claims and/or conclusions of the paper (regardless of whether the code and data are provided or not)?
    \item[] Answer: \answerYes{} 
    \item[] Justification: The paper specifies the benchmark axes, case tuple, construction pipeline, metrics, evaluation protocol, and appendices for prompts, calibration, and scoring, which together provide the information needed to reproduce the main results.
    \item[] Guidelines:
    \begin{itemize}
        \item The answer \answerNA{} means that the paper does not include experiments.
        \item If the paper includes experiments, a \answerNo{} answer to this question will not be perceived well by the reviewers: Making the paper reproducible is important, regardless of whether the code and data are provided or not.
        \item If the contribution is a dataset and\slash or model, the authors should describe the steps taken to make their results reproducible or verifiable. 
        \item Depending on the contribution, reproducibility can be accomplished in various ways. For example, if the contribution is a novel architecture, describing the architecture fully might suffice, or if the contribution is a specific model and empirical evaluation, it may be necessary to either make it possible for others to replicate the model with the same dataset, or provide access to the model. In general. releasing code and data is often one good way to accomplish this, but reproducibility can also be provided via detailed instructions for how to replicate the results, access to a hosted model (e.g., in the case of a large language model), releasing of a model checkpoint, or other means that are appropriate to the research performed.
        \item While NeurIPS does not require releasing code, the conference does require all submissions to provide some reasonable avenue for reproducibility, which may depend on the nature of the contribution. For example
        \begin{enumerate}
            \item If the contribution is primarily a new algorithm, the paper should make it clear how to reproduce that algorithm.
            \item If the contribution is primarily a new model architecture, the paper should describe the architecture clearly and fully.
            \item If the contribution is a new model (e.g., a large language model), then there should either be a way to access this model for reproducing the results or a way to reproduce the model (e.g., with an open-source dataset or instructions for how to construct the dataset).
            \item We recognize that reproducibility may be tricky in some cases, in which case authors are welcome to describe the particular way they provide for reproducibility. In the case of closed-source models, it may be that access to the model is limited in some way (e.g., to registered users), but it should be possible for other researchers to have some path to reproducing or verifying the results.
        \end{enumerate}
    \end{itemize}

\item {\bf Open access to data and code}
    \item[] Question: Does the paper provide open access to the data and code, with sufficient instructions to faithfully reproduce the main experimental results, as described in supplemental material?
    \item[] Answer: \answerYes{} 
    \item[] Justification: The abstract states that code, benchmark artifacts, and evaluation scripts are publicly released. 
    \item[] Guidelines:
    \begin{itemize}
        \item The answer \answerNA{} means that paper does not include experiments requiring code.
        \item Please see the NeurIPS code and data submission guidelines (\url{https://neurips.cc/public/guides/CodeSubmissionPolicy}) for more details.
        \item While we encourage the release of code and data, we understand that this might not be possible, so \answerNo{} is an acceptable answer. Papers cannot be rejected simply for not including code, unless this is central to the contribution (e.g., for a new open-source benchmark).
        \item The instructions should contain the exact command and environment needed to run to reproduce the results. See the NeurIPS code and data submission guidelines (\url{https://neurips.cc/public/guides/CodeSubmissionPolicy}) for more details.
        \item The authors should provide instructions on data access and preparation, including how to access the raw data, preprocessed data, intermediate data, and generated data, etc.
        \item The authors should provide scripts to reproduce all experimental results for the new proposed method and baselines. If only a subset of experiments are reproducible, they should state which ones are omitted from the script and why.
        \item At submission time, to preserve anonymity, the authors should release anonymized versions (if applicable).
        \item Providing as much information as possible in supplemental material (appended to the paper) is recommended, but including URLs to data and code is permitted.
    \end{itemize}

\item {\bf Experimental setting/details}
    \item[] Question: Does the paper specify all the training and test details (e.g., data splits, hyperparameters, how they were chosen, type of optimizer) necessary to understand the results?
    \item[] Answer: \answerYes{} 
    \item[] Justification: The paper describes the evaluated models, memory substrates, paired clean/poisoned protocol, defense families, metrics, and points to appendices for decoding settings, prompts, thresholds, and calibration details.
    \item[] Guidelines:
    \begin{itemize}
        \item The answer \answerNA{} means that the paper does not include experiments.
        \item The experimental setting should be presented in the core of the paper to a level of detail that is necessary to appreciate the results and make sense of them.
        \item The full details can be provided either with the code, in appendix, or as supplemental material.
    \end{itemize}

\item {\bf Experiment statistical significance}
    \item[] Question: Does the paper report error bars suitably and correctly defined or other appropriate information about the statistical significance of the experiments?
    \item[] Answer: \answerYes{} 
    \item[] Justification: We report statistical uncertainty for the main experimental results using standard deviations computed across the five runs in the main tables; in addition, figures that aggregate results across models include error bars showing the standard deviation over the ten evaluated models.
    \item[] Guidelines:
    \begin{itemize}
        \item The answer \answerNA{} means that the paper does not include experiments.
        \item The authors should answer \answerYes{} if the results are accompanied by error bars, confidence intervals, or statistical significance tests, at least for the experiments that support the main claims of the paper.
        \item The factors of variability that the error bars are capturing should be clearly stated (for example, train/test split, initialization, random drawing of some parameter, or overall run with given experimental conditions).
        \item The method for calculating the error bars should be explained (closed form formula, call to a library function, bootstrap, etc.)
        \item The assumptions made should be given (e.g., Normally distributed errors).
        \item It should be clear whether the error bar is the standard deviation or the standard error of the mean.
        \item It is OK to report 1-sigma error bars, but one should state it. The authors should preferably report a 2-sigma error bar than state that they have a 96\% CI, if the hypothesis of Normality of errors is not verified.
        \item For asymmetric distributions, the authors should be careful not to show in tables or figures symmetric error bars that would yield results that are out of range (e.g., negative error rates).
        \item If error bars are reported in tables or plots, the authors should explain in the text how they were calculated and reference the corresponding figures or tables in the text.
    \end{itemize}

\item {\bf Experiments compute resources}
    \item[] Question: For each experiment, does the paper provide sufficient information on the computer resources (type of compute workers, memory, time of execution) needed to reproduce the experiments?
    \item[] Answer: \answerYes{}{} 
    \item[] Justification: The experimental compute resources, including hardware setup, GPU runtime, and API token usage, are reported in the appendix.
    \item[] Guidelines:
    \begin{itemize}
        \item The answer \answerNA{} means that the paper does not include experiments.
        \item The paper should indicate the type of compute workers CPU or GPU, internal cluster, or cloud provider, including relevant memory and storage.
        \item The paper should provide the amount of compute required for each of the individual experimental runs as well as estimate the total compute. 
        \item The paper should disclose whether the full research project required more compute than the experiments reported in the paper (e.g., preliminary or failed experiments that didn't make it into the paper). 
    \end{itemize}
    
\item {\bf Code of ethics}
    \item[] Question: Does the research conducted in the paper conform, in every respect, with the NeurIPS Code of Ethics \url{https://neurips.cc/public/EthicsGuidelines}?
    \item[] Answer: \answerYes{}{}{} 
    \item[] Justification: To the best of our knowledge, the research conforms to the NeurIPS Code of Ethics. The work studies security risks in agent memory with the goal of improving robustness and understanding defense failures.
    \item[] Guidelines:
    \begin{itemize}
        \item The answer \answerNA{} means that the authors have not reviewed the NeurIPS Code of Ethics.
        \item If the authors answer \answerNo, they should explain the special circumstances that require a deviation from the Code of Ethics.
        \item The authors should make sure to preserve anonymity (e.g., if there is a special consideration due to laws or regulations in their jurisdiction).
    \end{itemize}

\item {\bf Broader impacts}
    \item[] Question: Does the paper discuss both potential positive societal impacts and negative societal impacts of the work performed?
    \item[] Answer: \answerYes{} 
    \item[] Justification: The paper discusses both positive societal impacts, including improved robustness and safer deployment of memory enabled agents, and negative impacts, including dual-use risks and the possibility that overly aggressive defenses may harm benign personalization or utility.
    \item[] Guidelines:
    \begin{itemize}
        \item The answer \answerNA{} means that there is no societal impact of the work performed.
        \item If the authors answer \answerNA{} or \answerNo, they should explain why their work has no societal impact or why the paper does not address societal impact.
        \item Examples of negative societal impacts include potential malicious or unintended uses (e.g., disinformation, generating fake profiles, surveillance), fairness considerations (e.g., deployment of technologies that could make decisions that unfairly impact specific groups), privacy considerations, and security considerations.
        \item The conference expects that many papers will be foundational research and not tied to particular applications, let alone deployments. However, if there is a direct path to any negative applications, the authors should point it out. For example, it is legitimate to point out that an improvement in the quality of generative models could be used to generate Deepfakes for disinformation. On the other hand, it is not needed to point out that a generic algorithm for optimizing neural networks could enable people to train models that generate Deepfakes faster.
        \item The authors should consider possible harms that could arise when the technology is being used as intended and functioning correctly, harms that could arise when the technology is being used as intended but gives incorrect results, and harms following from (intentional or unintentional) misuse of the technology.
        \item If there are negative societal impacts, the authors could also discuss possible mitigation strategies (e.g., gated release of models, providing defenses in addition to attacks, mechanisms for monitoring misuse, mechanisms to monitor how a system learns from feedback over time, improving the efficiency and accessibility of ML).
    \end{itemize}
    
\item {\bf Safeguards}
    \item[] Question: Does the paper describe safeguards that have been put in place for responsible release of data or models that have a high risk for misuse (e.g., pre-trained language models, image generators, or scraped datasets)?
    \item[] Answer: \answerNA{} 
    \item[] Justification: The paper does not release a pretrained high-risk model or a scraped sensitive dataset. It releases a text benchmark and evaluation artifacts.
    \item[] Guidelines:
    \begin{itemize}
        \item The answer \answerNA{} means that the paper poses no such risks.
        \item Released models that have a high risk for misuse or dual-use should be released with necessary safeguards to allow for controlled use of the model, for example by requiring that users adhere to usage guidelines or restrictions to access the model or implementing safety filters. 
        \item Datasets that have been scraped from the Internet could pose safety risks. The authors should describe how they avoided releasing unsafe images.
        \item We recognize that providing effective safeguards is challenging, and many papers do not require this, but we encourage authors to take this into account and make a best faith effort.
    \end{itemize}

\item {\bf Licenses for existing assets}
    \item[] Question: Are the creators or original owners of assets (e.g., code, data, models), used in the paper, properly credited and are the license and terms of use explicitly mentioned and properly respected?
    \item[] Answer: \answerYes{} 
    \item[] Justification: We properly credit the existing assets used in this work, and their versions, licenses, and terms of use are listed in the release materials like croissant file and the corresponding assets are cited in the paper.
    \item[] Guidelines:
    \begin{itemize}
        \item The answer \answerNA{} means that the paper does not use existing assets.
        \item The authors should cite the original paper that produced the code package or dataset.
        \item The authors should state which version of the asset is used and, if possible, include a URL.
        \item The name of the license (e.g., CC-BY 4.0) should be included for each asset.
        \item For scraped data from a particular source (e.g., website), the copyright and terms of service of that source should be provided.
        \item If assets are released, the license, copyright information, and terms of use in the package should be provided. For popular datasets, \url{paperswithcode.com/datasets} has curated licenses for some datasets. Their licensing guide can help determine the license of a dataset.
        \item For existing datasets that are re-packaged, both the original license and the license of the derived asset (if it has changed) should be provided.
        \item If this information is not available online, the authors are encouraged to reach out to the asset's creators.
    \end{itemize}

\item {\bf New assets}
    \item[] Question: Are new assets introduced in the paper well documented and is the documentation provided alongside the assets?
    \item[] Answer: \answerYes{} 
    \item[] Justification: The paper introduces a new benchmark and evaluation artifacts, and documents their scope, schema, construction pipeline, and evaluation protocol. The release package also includes structured documentation.
    \item[] Guidelines:
    \begin{itemize}
        \item The answer \answerNA{} means that the paper does not release new assets.
        \item Researchers should communicate the details of the dataset\slash code\slash model as part of their submissions via structured templates. This includes details about training, license, limitations, etc. 
        \item The paper should discuss whether and how consent was obtained from people whose asset is used.
        \item At submission time, remember to anonymize your assets (if applicable). You can either create an anonymized URL or include an anonymized zip file.
    \end{itemize}

\item {\bf Crowdsourcing and research with human subjects}
    \item[] Question: For crowdsourcing experiments and research with human subjects, does the paper include the full text of instructions given to participants and screenshots, if applicable, as well as details about compensation (if any)? 
    \item[] Answer: \answerNA{} 
    \item[] Justification: The paper does not report crowdsourcing experiments or human-subject studies as part of its main contribution.
    \item[] Guidelines:
    \begin{itemize}
        \item The answer \answerNA{} means that the paper does not involve crowdsourcing nor research with human subjects.
        \item Including this information in the supplemental material is fine, but if the main contribution of the paper involves human subjects, then as much detail as possible should be included in the main paper. 
        \item According to the NeurIPS Code of Ethics, workers involved in data collection, curation, or other labor should be paid at least the minimum wage in the country of the data collector. 
    \end{itemize}

\item {\bf Institutional review board (IRB) approvals or equivalent for research with human subjects}
    \item[] Question: Does the paper describe potential risks incurred by study participants, whether such risks were disclosed to the subjects, and whether Institutional Review Board (IRB) approvals (or an equivalent approval/review based on the requirements of your country or institution) were obtained?
    \item[] Answer: \answerNA{} 
    \item[] Justification: The work does not appear to involve human-subject experimentation requiring IRB approval.
    \item[] Guidelines:
    \begin{itemize}
        \item The answer \answerNA{} means that the paper does not involve crowdsourcing nor research with human subjects.
        \item Depending on the country in which research is conducted, IRB approval (or equivalent) may be required for any human subjects research. If you obtained IRB approval, you should clearly state this in the paper. 
        \item We recognize that the procedures for this may vary significantly between institutions and locations, and we expect authors to adhere to the NeurIPS Code of Ethics and the guidelines for their institution. 
        \item For initial submissions, do not include any information that would break anonymity (if applicable), such as the institution conducting the review.
    \end{itemize}

\item {\bf Declaration of LLM usage}
    \item[] Question: Does the paper describe the usage of LLMs if it is an important, original, or non-standard component of the core methods in this research? Note that if the LLM is used only for writing, editing, or formatting purposes and does \emph{not} impact the core methodology, scientific rigor, or originality of the research, declaration is not required.
    \item[] Answer: \answerYes{} 
    \item[] Justification: LLMs are central to the experimental setup: the paper evaluates multiple open- and closed-weight LLM families and also describes model-based judge prompts in the evaluation pipeline and defense baselines.
    \item[] Guidelines:
    \begin{itemize}
        \item The answer \answerNA{} means that the core method development in this research does not involve LLMs as any important, original, or non-standard components.
        \item Please refer to our LLM policy in the NeurIPS handbook for what should or should not be described.
    \end{itemize}

\end{enumerate}

\end{document}